\documentclass[a4paper,11pt]{article}

\pdfoutput=1

\usepackage{jcappub}
\usepackage{amsmath}
\usepackage{amsfonts}
\usepackage{amssymb}
\usepackage{mathtools}
\usepackage{graphics}
\usepackage{citesort}
\usepackage{graphicx}

\usepackage[utf8]{inputenc}

\usepackage{hyperref}

\newcommand{\eps}{\epsilon}

\newcommand{\e}[1]{\times10^{#1}}

\title{The Bright and Choked Gamma-Ray Burst Contribution to the IceCube and ANTARES Low-Energy Excess}

\author[a]{Peter B.~Denton}
\emailAdd{denton@nbi.ku.dk}
\author[a,b]{and Irene Tamborra}
\emailAdd{tamborra@nbi.ku.dk}

\affiliation[a]{Niels Bohr International Academy, Niels Bohr Institute, University of Copenhagen, Blegdamsvej 17, 2100, Copenhagen, Denmark}
\affiliation[b]{DARK, Niels Bohr Institute, University of Copenhagen, Juliane Maries Vej 30, 2100, Copenhagen, Denmark}

\abstract{
The increasing statistics of the high-energy neutrino flux observed by the IceCube Observatory points towards an excess of events above the atmospheric neutrino background in the 30--400~TeV energy range.
Such an excess is compatible with the findings of the ANTARES Telescope and it would naturally imply the possibility that more than one source class contributes to the observed flux. 
Electromagnetically hidden sources have been invoked to interpret this excess of events at low energies. By adopting a unified model for the electromagnetically bright and choked gamma-ray bursts and taking into account particle acceleration at the internal and collimation shock radii, we discuss whether bright and choked bursts are viable candidates.
Our findings suggest that, although producing a copious neutrino flux, choked and bright astrophysical jets cannot be the dominant sources of the excess of neutrino events.
A fine tuning of the model parameters or distinct scenarios for choked jets should be invoked in order to explain the low-energy neutrino data of IceCube and ANTARES.}

\begin{document}

\maketitle

\section{Introduction}\label{sec:introduction}
The IceCube telescope detected astrophysical neutrinos with the highest energies ever observed~\cite{Aartsen:2013bka,Aartsen:2013jdh,Aartsen:2014gkd,Aartsen:2014muf,Aartsen:2015knd,Aartsen:2015rwa,Aartsen:2015zva,Aartsen:2017mau}.
The sources of these neutrinos remain to be discovered.
Current data show evidence for a major extragalactic component contributing to the observed flux~\cite{Palladino:2016zoe,Denton:2017csz,Aartsen:2017ujz}.
A number of astrophysical sources has been discussed as possibly being at the origin of the observed flux: starburst galaxies, clusters of galaxies, active galactic nuclei, low-power astrophysical jets, and tidal disruption events~\cite{Meszaros:2015krr,Waxman:2015ues,Murase:2015ndr,Ahlers:2015lln,Anchordoqui:2013dnh,Dai:2016gtz,Senno:2016bso,Lunardini:2016xwi}.
Noticeably, it remains to be clarified whether the observed flux is the result of the superposition of fluxes coming from different classes of sources, or from a single class.

\begin{figure}
\centering
\includegraphics[width=5in]{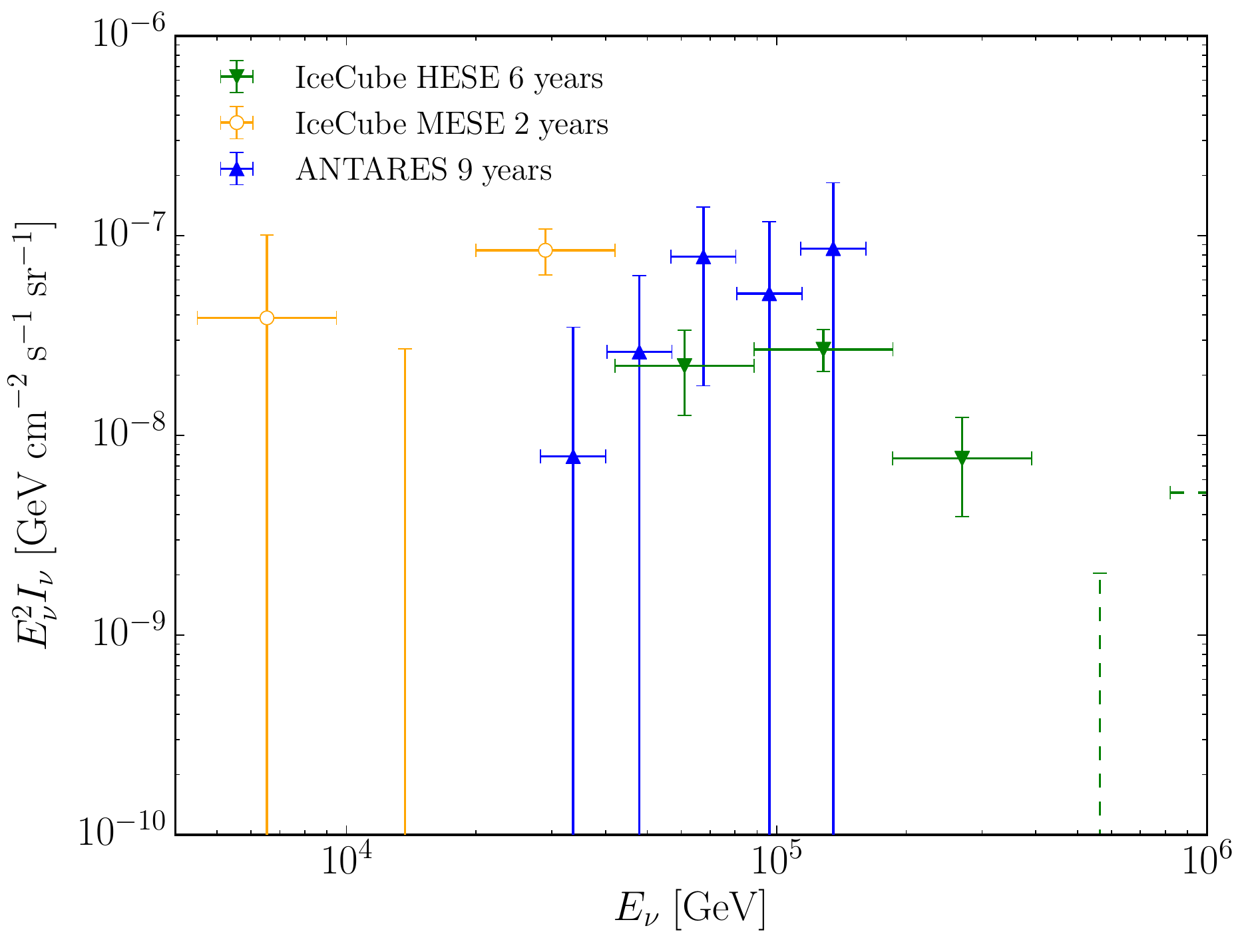}
\caption{IceCube and ANTARES data sets in the energy region of interest. The IceCube High Energy Starting Event (HESE) set refers to 6 years of data taking as from Ref.~\cite{Aartsen:2017mau} and the IceCube Medium Energy Starting Event (MESE) set refers to 2 years of data taking~\cite{Aartsen:2014muf}. The ANTARES 9-years event rate above atmospheric backgrounds \cite{Albert:2017nsd} has been folded with the detector effective area \cite{AdrianMartinez:2012rp} (see Sec.~\ref{sec:excess} for more details).
We take only the three lowest bins from the MESE data set as the higher energy bins contain largely the same events as the HESE data which now has more statistics.}
\label{fig:Inu1}
\end{figure}

Mounting evidence points towards the possibility that fitting a single power law to the observed astrophysical flux is disfavored, suggesting that more than one class of sources is contributing to the flux \cite{Chen:2014gxa,Chianese:2016opp,Anchordoqui:2016ewn,Palladino:2017qda,Palladino:2018evm}.
In particular, data recently presented by the ANTARES~\cite{Albert:2017nsd} and IceCube~\cite{Aartsen:2014muf,Aartsen:2015zva,Aartsen:2017mau} Collaborations independently point towards an excess of neutrino events in the energy range between 30 and 400 TeV as shown in Fig.~\ref{fig:Inu1} (see also e.g.~Fig.~3 of Ref.~\cite{Chianese:2017jfa})\footnote{The statistical significance of an excess from ANTARES data alone is weak. Taking optimistic values on uncertainties we find $\chi^2=0.44$ evidence for an excess above the expected astrophysical signal.}.
In this paper, we will take this hint of an excess at low energies or break in the energy spectrum seriously and investigate its origin.

Since none of the above proposed sources seems to be able to fully explain the observed flux, sources that are electromagnetically dim have been proposed as a viable alternative option, especially for what concerns the low-energy tail of the neutrino spectrum~\cite{Murase:2015xka,Ahlers:2014ioa}. Invoking the contribution from hidden or low-power sources could also alleviate conflict with the electromagnetic counterparts observed by {\it Fermi}~\cite{Ando:2015bva,Bechtol:2015uqb} and may be testable by stacking searches in the near future~\cite{Murase:2016gly,Mertsch:2016hcd}. In particular, ``choked'' gamma-ray bursts (GRBs)~\cite{Meszaros:2001ms,Ando:2005xi,Horiuchi:2007xi,Razzaque:2003uw,Razzaque:2003uv,Razzaque:2002kb}, i.e., GRBs that are not electromagnetically bright but emit neutrinos, have been proposed as candidate sources of the IceCube neutrinos~\cite{Tamborra:2015fzv,Senno:2015tsn,Senno:2017vtd,Murase:2013ffa}. 

Recently, Ref.~\cite{Denton:2017jwk} presented a general model aiming to unify the neutrino production in electromagnetically bright and choked GRBs. The model parameters were tuned on the electromagnetically bright jets and extrapolated to the choked jets by assuming that choked jets are harbored in massive stars similarly to electromagnetically luminous jets. By adopting the six-year high-energy starting events sample from IceCube~\cite{Aartsen:2017mau} as an upper limit to the neutrino flux produced by luminous and choked jets, it was found that at most $1\%$ of all core-collapse supernovae can harbor astrophysical jets and the majority of those jets are choked. 

In this work, we rely on the ``advanced'' GRB model of Ref.~\cite{Denton:2017jwk} that employs a realistic distribution of the Lorentz boost factor within the jet.
(A similar model was more recently discussed in Ref.~\cite{Biehl:2017qen} in the context of short GRBs.)
For the first time, we here discuss particle acceleration at the internal (IS) and collimation (CS) shock radii and explore whether bright and choked jets can lead to a sizable neutrino production in the low-energy range of interest, as postulated formerly.
Our findings are then investigated in the context of the excess of neutrino events observed by IceCube and ANTARES in the 30--400 TeV window. 

This manuscript is organized as follows. In Sec.~\ref{sec:nuflux}, the neutrino emission within the advanced GRB model is introduced as well as the various cooling processes affecting the neutrino spectrum at the internal and collimation shocks. In Sec.~\ref{sec:excess}, we investigate whether the observed neutrino flux in the energy window between 30 and 400 TeV can be interpreted in terms of electromagnetically bright and choked GRBs. In Sec.~\ref{sec:conclusions}, a discussion on our findings and conclusions are reported. 

\section{Neutrino emission from bright and choked gamma-ray bursts}\label{sec:nuflux}
In this Section, the ``advanced'' GRB model is introduced. We also overview the properties of cooling processes mainly determining the observed flux of neutrinos and affecting protons and secondary particles.

\subsection{Properties of the astrophysical jet}
\label{ssec:jet properties}
\begin{figure}
\centering
\includegraphics[height=3in]{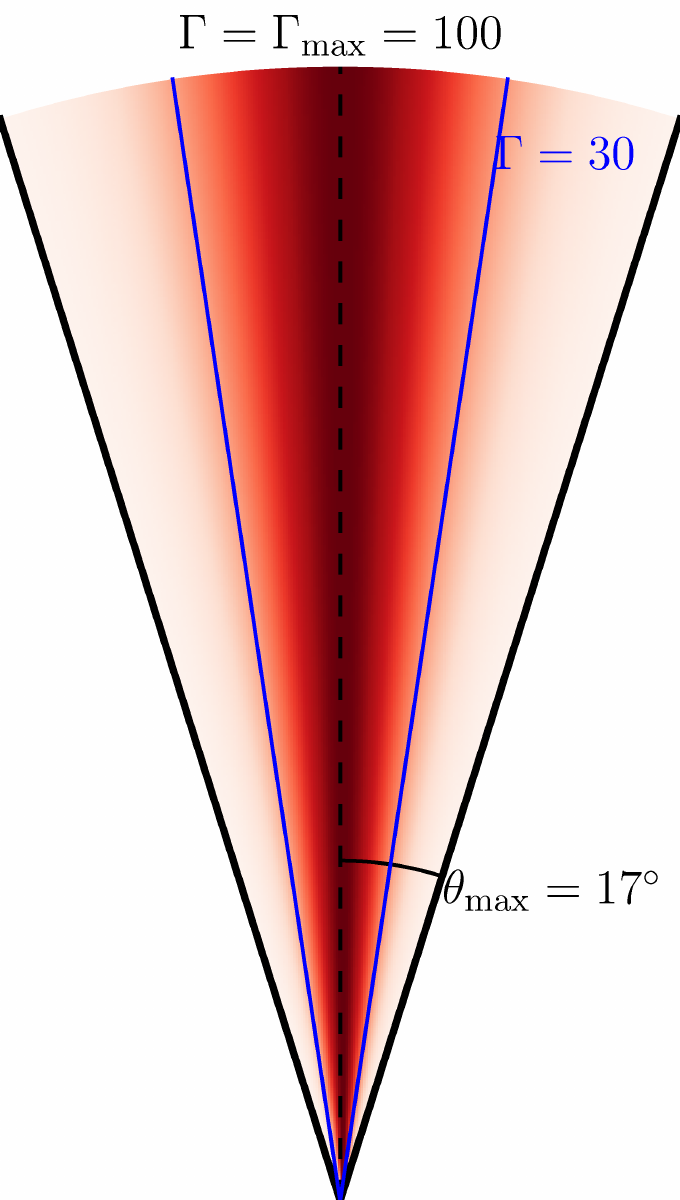}
\caption{The structure of the jet within our ``advanced'' GRB model.
The Lorentz boost factor $\Gamma$ is shown in red and is maximum along the axis of the jet. The color scaling indicates the distribution in $\Gamma$ throughout the jet and it has been derived for an astrophysical jet with $\Gamma_{\mathrm{max}}=100$ and $\theta_{\max}=17^\circ$.}
\label{fig:schematic advanced}
\end{figure}

We parametrize the astrophysical jet in terms of the amount of kinetic energy in the jet $\tilde E_j$, maximum bulk Lorentz factor of the jet $\Gamma_{\max}$, and electron (magnetic) energy fraction $\eps_e$ ($\eps_B$). We then follow the ``advanced" GRB model of Ref.~\cite{Denton:2017jwk} wherein the Lorentz factor varies throughout the jet.
Along the jet axis $\Gamma=\Gamma_{\max}$ and it falls off from there and is given by
\begin{equation}
\Gamma(\theta)=\Gamma_{\max}\exp[\kappa(\cos\theta-1)]\,,
\end{equation}
where the concentration is related to the standard deviation $\kappa\approx1/\sigma^2$ as shown in Fig.~\ref{fig:schematic advanced} (the color scaling stands for the $\Gamma$ distribution throughout the jet).
The standard deviation is set to $\sigma=1/\sqrt{\Gamma_{\max}}$ motivated by a random walk of repeated shocks.
The maximum angle of the jet is defined by $\Gamma(\theta_{\max})=1$ and is
\begin{equation}
\theta_{\max}=\cos^{-1}\left(1-\frac{\ln\Gamma_{\max}}{\Gamma_{\max}}\right)\,.
\end{equation}

The jet volume element is given by $V'=\Omega_j\tilde r_j^2c\tilde t_j\Gamma$, and the solid angle for both jets is,
\begin{equation}
\Omega_j=4\pi(1-\cos\theta_{\max})\approx2\pi\theta_{\max}^2\,.
\end{equation}
We distinguish among the three relevant reference frames: $X$ - earth, $\tilde X$ - GRB, $X'$ - jet\footnote{
Energies in each frame are related by $\tilde E=(1+z)E$, and $\tilde E=\Gamma E'$.
Times are related by $t=(1+z)\tilde t$, and $t'=\Gamma\tilde t$.
Luminosities are related by $\tilde L=(1+z)^2L$.}.

\begin{figure}
\centering
\includegraphics[width=0.32\textwidth]{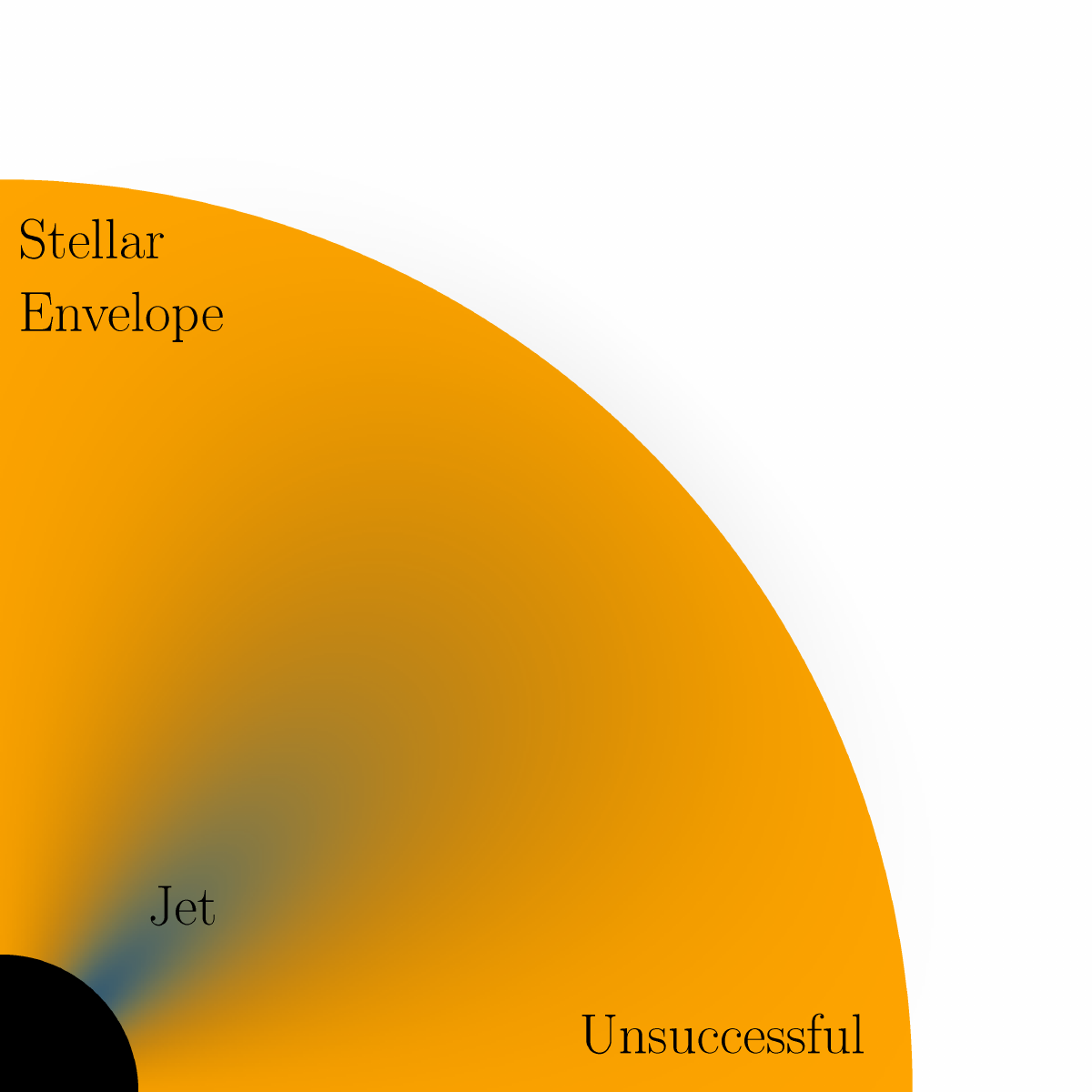}
\includegraphics[width=0.32\textwidth]{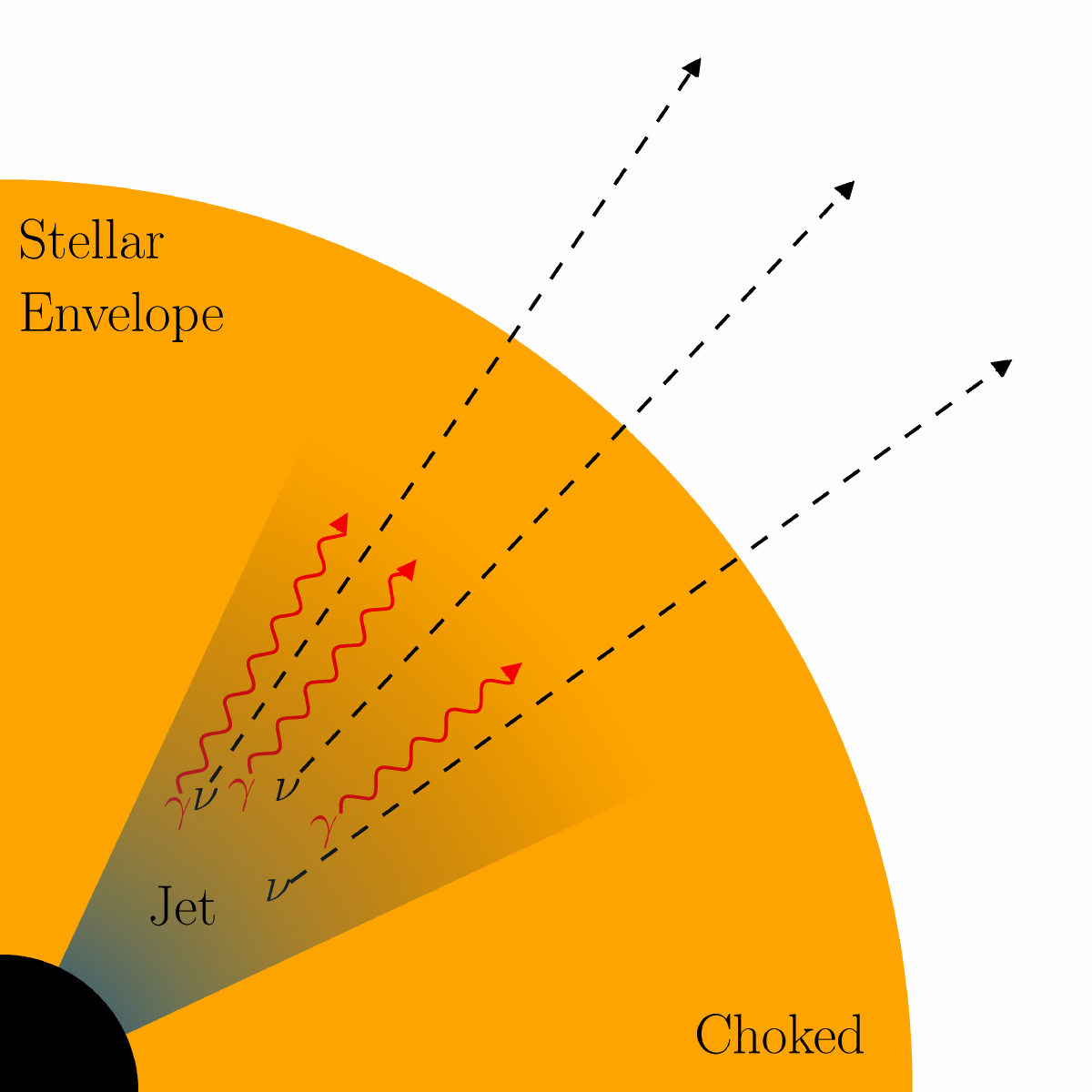}
\includegraphics[width=0.32\textwidth]{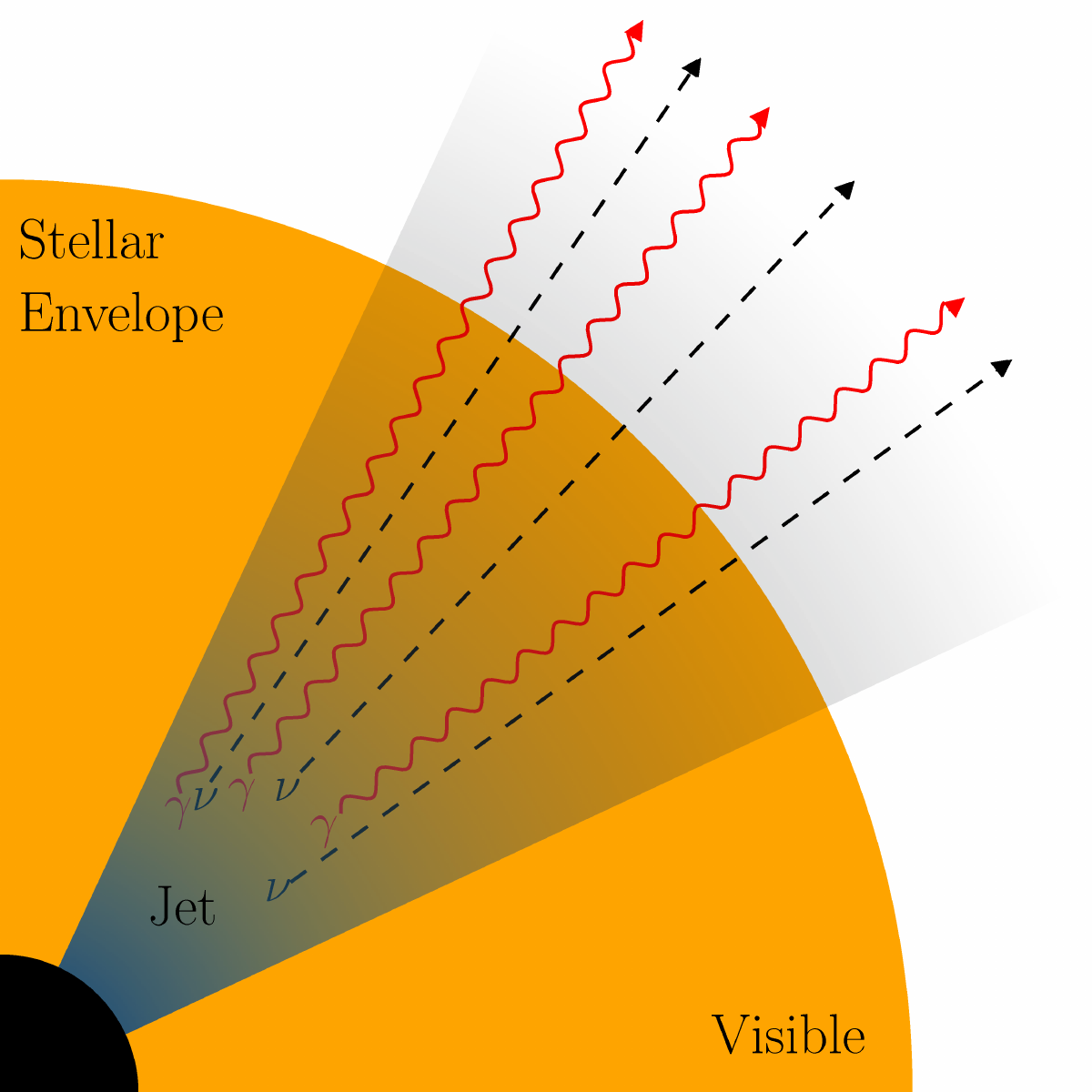}
\caption{Three classifications of jet propagating within the stellar envelope are possible.
If the jet is optically thick near the acceleration region, the jet will not accelerate particles to high energies, this is known as an unsuccessful jet (left panel).
If the jet successfully accelerates particles but does not have enough energy to escape the stellar envelope, the jet will be choked (middle panel): high energy neutrinos will escape, but no electromagnetic radiation.
If the jet successfully accelerates particles and the jet escapes the stellar envelope, then it is visible (right panel) and both high energy neutrinos and electromagnetic radiation escape. Particle production and acceleration occurs in proximity of the internal or collimation shock radii.}
\label{fig:schematic}
\end{figure}

The volume element and the jet energy together determine the density inside the jet and its Thomson optical depth $\tau'_T$.
If the optical depth is large ($\tau'_T\gtrsim 1$) then efficient particle acceleration cannot happen and the jet fails to produce any high energy particles \cite{Murase:2013ffa,Denton:2017jwk}.
We call these jets unsuccessful (see left panel of Fig.~\ref{fig:schematic}). If $\tau'_T\lesssim 1$, then the jet is successful in accelerating particles; in the latter case, the jet can be choked (middle panel of Fig.~\ref{fig:schematic}) or visible (right panel of Fig.~\ref{fig:schematic}) if only neutrinos or neutrinos and photons are able to escape the jet respectively.

In order to be consistent with electromagnetic observations, for the jet duration $t_j$, we use a power law relation to describe long-duration GRBs such that jets with higher $\Gamma$'s are shorter than GRBs with lower $\Gamma$'s.
Then, $\tilde t_j\propto\frac1\Gamma$ is normalized to $\tilde t_j=10$ s at $\Gamma=300$ to match the trend shown by observations \cite{Gehrels:2013xd,Lu:2017toj}.

The photon spectrum of HL-GRBs is observed to be non-thermal and is well described by the Band spectrum \cite{Band:1993eg} with the break energy related to the jet energy by the Amati and Yonetoku relations \cite{Amati:2002ny,Yonetoku:2003gi}, each of which is the result of fits to the data.
As described in Ref.~\cite{Denton:2017jwk}, we assume that the non-thermal photon spectrum which applies for high-luminosity (HL-) GRBs can be extended to all GRBs.
Note that if the photon spectrum is thermal, then efficient particle acceleration is impossible \cite{Murase:2013ffa}; we consider these cases to be unsuccessful at accelerating any high energy particles and the flux of high energy neutrinos is then zero.

Finally, we include all cooling processes within the jet: synchrotron, inverse Compton, Bethe-Heitler, and scattering off both protons and photons using measured cross sections as from Ref.~\cite{Olive:2016xmw}, and take $\eps_e=\eps_B=0.1$.
We direct the interested reader to Ref.~\cite{Denton:2017jwk} for more details.

\subsection{Particle acceleration at collimation and internal shock radii}
\label{sec:radius}
The IS model is empirically characterized by the variability time scale under the premise that particle production cannot be happening at radii larger than $r_{\rm IS}$ as the regions would be causally disconnected.
As such, the lower limit for the IS radius is given by the variability time whose scaling law is established on the grounds of observations.

The theoretically derived CS model instead includes a cocoon around the jet of shocked material that pushes back on the jet and collimates it \cite{Bromberg:2011fg}.
In other words, the CS radius is mainly depending on model parameters not directly measurable.

It might be possible that the CS radius coincides with the IS radius.
That is, if the radius derived by variability time measurements is comparable with the theoretical prediction for the location of the CS radius, then these separate models would be equivalent.
In the more general event that they are different, it is possible that there is significant particle acceleration at both radii, if $r_{\rm CS}>r_{\rm IS}$.
If $r_{\rm IS}>r_{\rm CS}$, then there cannot be a significant contribution to particle acceleration at the CS because that would introduce a new smaller variability time scale that would contradict observations.
We note that in some models it may be possible to somewhat avoid this constraint, for example by invoking effects due to magnetic reconnection~\cite{Thompson:1994zh} or acceleration at other radii \cite{Biehl:2017zlw}.
However, the neutrino production in these models peaks at $E_\nu\gtrsim1$ PeV~\cite{Zhang:2010jt,Zhang:2012qy}, and would not improve the fit to the low energy data considered here.

The neutrino production at both radii was first estimated in Ref.~\cite{Murase:2013ffa}, which concluded that the neutrino production from the CS peaks at lower energies than in the IS model, but unless the jet is extremely relativistic, then $r_{\rm IS} < r_{\rm CS}$ for typical Wolf-Rayet progenitors.
Given the sophistications introduced in the advanced GRB model and since we are focusing on the low energy tail of the neutrino spectrum, we will first investigate whether there is any contribution to the neutrino intensity from acceleration occurring at the CS radius in the advanced GRB model. 
In the following, we are going to compare these two radii within the advanced GRB model, in order to test at which radius particle production can occur.

The IS radius is defined as $\tilde r_{\rm IS}=2c\tilde t_v\Gamma^2$.
In the advanced GRB model, the jet variability time is taken from an empirical fit to HL-GRBs \cite{Sonbas:2014jya} and a maximum to cap the variability time for low-luminosity (LL-) GRBs~\cite{Gupta:2006jm,Campana:2006qe,Denton:2017jwk}:
$\tilde t_v=\min(2.8\e9\Gamma^{-4.05},100)\,\text s$.
Therefore the IS radius is at
\begin{equation}
\frac{\tilde r_{\rm IS}}{\rm cm}=2.9\e{16}\times
\begin{cases}
\left(\frac\Gamma{69}\right)^2&\Gamma<69\\
\left(\frac\Gamma{69}\right)^{-2.05}\quad&\Gamma>69
\end{cases}\,.
\end{equation}
Note that the strong scaling of the variability time from observations results in considerable differences in this definition of the IS radius from that of the model presented in Ref.~\cite{Murase:2013ffa}.

From Refs.~\cite{Bromberg:2011fg,Murase:2013ffa}, the CS radius is at
\begin{equation}
\frac{\tilde r_{\rm CS}}{\rm cm}=2.4\e9\,\tilde t_{j,1}^{8/5}\tilde L_{\mathrm{iso},52}^{6/5}\left(\frac{\theta_{\max}}{0.2}\right)^{-4/5}\left(\frac{M_*}{20M_\odot}\right)^{-6/5}R_{*,11}^{3/5}\,,
\end{equation}
where we used the standard notation, $Q_x=Q/10^x$ in cgs units.
The radius $R_\star\sim R_\odot$ is the progenitor radius of a typical Wolf-Rayet star and $\tilde L_{\mathrm{iso}}\simeq (4 \pi \tilde E_j)/(\tilde t_j \Omega_j)$ is the isotropic equivalent luminosity.

\begin{figure}
\centering
\includegraphics[width=5in]{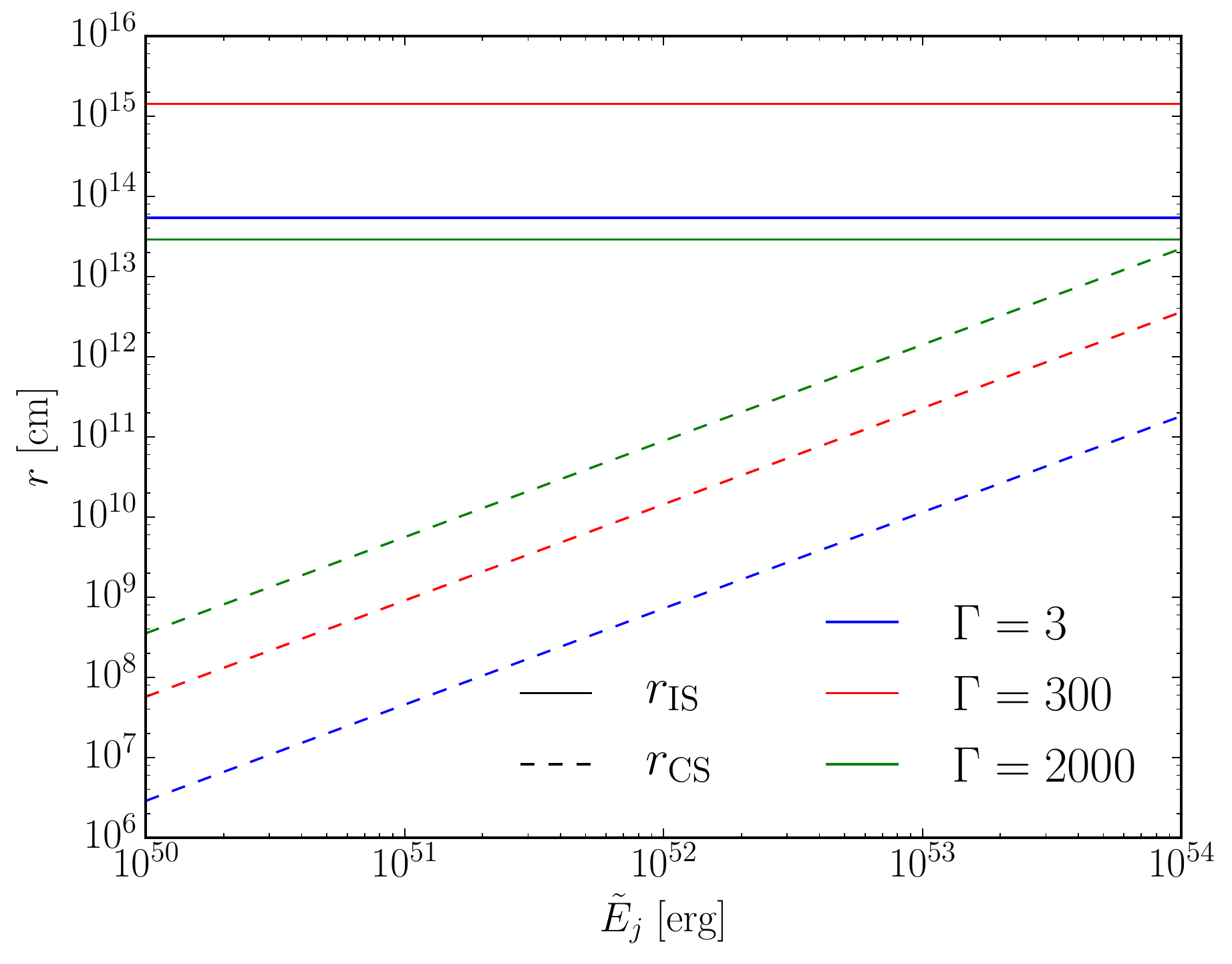}
\caption{The radius of acceleration due to internal shock (IS) and collimation shock (CS) at different Lorentz boost factors.
The IS radii are the solid curves while the CS radii are the dashed lines.
The IS radii are always larger than CS radii except for extreme cases such as $\Gamma_{\max}\gtrsim2000$ and $\tilde E_j\gtrsim10^{54}$ erg.}
\label{fig:rIS_rCS}
\end{figure}

By comparing the IS and CS radii within the advanced GRB model for a typical Wolf-Rayet progenitor with mass $20\ M_\odot$, we see that for nearly all relevant values of the parameters of the advanced GRB model, $r_{\rm IS}>r_{\rm CS}$ (note that this trend is opposite than the findings of Ref.~\cite{Murase:2013ffa} because of the differences in the values assumed for $\tilde t_v$ and the sophistications introduced in the advanced GRB model). This trend is shown in Fig.~\ref{fig:rIS_rCS} where the IS and CS radii are plotted as a function of jet energy for two representative values of $\Gamma$. 
We conclude that there can be no acceleration at $r_{\rm CS}$ as that would introduce a smaller variability time scale in contradiction with the observations of HL- and LL-GRBs on which the advanced model is tuned.

Only for fairly extreme jets with $\Gamma_{\max}\gtrsim2000$ and $\tilde E_j\gtrsim10^{54}$ erg do we find $r_{\rm CS}\sim r_{\rm IS}$.
Since these are parameters at the tail end of the GRB distribution in both $\Gamma_{\max}$ and $\tilde E_j$ and because including particle acceleration at two similar radii does not change the observed spectrum, we will consider fully efficient particle acceleration at the IS radius in the following.

It is also possible that a small fraction of the total acceleration occurs at the CS radius and the consequently smaller variability time is not measured.
For the sake of simplicity, in the following we will assume that acceleration is fully efficient at the IS radius.
However, since there are not measurements of choked jets, it is possible that acceleration in those jets is different.
For completeness, we will also discuss the possibility that particles are accelerated at the CS radius in choked jets and estimate the correspondent neutrino intensity.

\subsection{Conditions for obtaining electromagnetically bright and choked jets}
If the jet head does not escape the stellar envelope, then the jet is considered to be choked (middle panel of Fig.~\ref{fig:schematic}).
Choked jets experience high energy particle acceleration in the vicinity of the acceleration region given by the IS radius (see Sec.~\ref{sec:radius}), but the jet does not escape the stellar envelope.
In this case, the high energy photons rapidly lose energy due to pair production and are not observed at the Earth, but the high energy neutrinos will escape and can still be observed.

\begin{figure}
\centering
\includegraphics[width=5in]{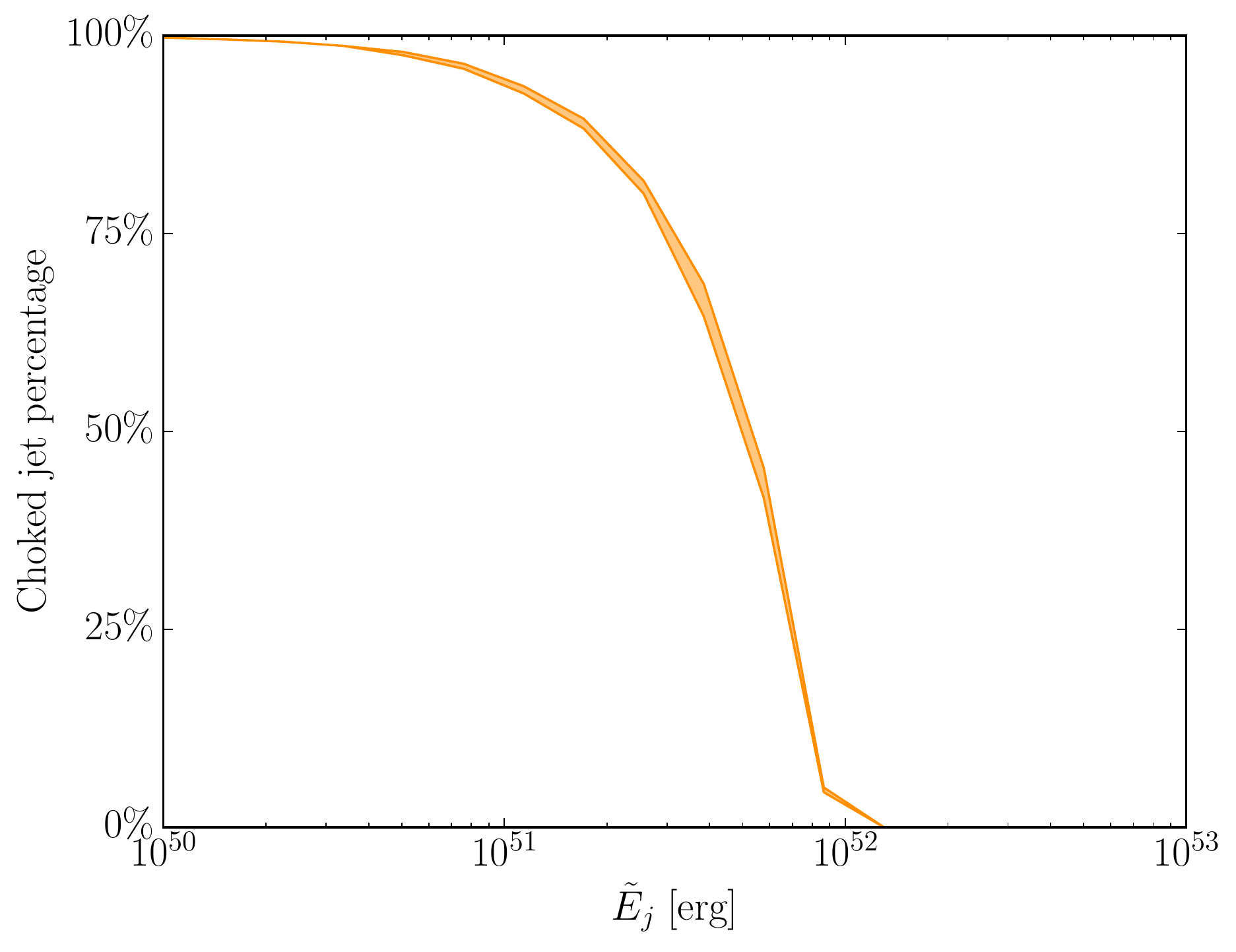}
\caption{The percentage of successful jets out of the distribution of $\Gamma_{\max}$ that are choked.
As the jet energy increases, jets become increasingly likely to be successful.
The band represents the small additional variation due to $\zeta_{\rm SN}\in[0.01,1]$ which affects the distribution of how relativistic the jets are.}
\label{fig:choked}
\end{figure}

The jet head ($h$) radius is at~\cite{Bromberg:2011fg} 
\begin{equation}
\tilde r_h=5.4\e{10}{\rm\ cm\ }\tilde t_{j,1}^{6/5}\tilde L_{{\rm iso},52}^{2/5}
\times\left(\frac{\theta_{\max}}{0.2}\right)^{-8/5}\left(\frac{M_*}{20\ M_\odot}\right)^{-2/5}R_{*,11}^{1/5}\,.
\label{eq:r_h}
\end{equation}
We then compare $\tilde r_h$ to a typical stellar envelope from a Wolf-Rayet star.
Figure~\ref{fig:choked} shows the fraction of successful jets out of the distribution of $\Gamma_{\max}$ that will be choked in the region of interest.
One can see that, as the jet energy increases, the probability that the jet successfully breaks through the stellar envelope increases and the fraction of choked jets becomes smaller.
In addition, the band represents the variation from $\zeta_{\rm SN}\in[0.01,1]$ which has very little influence on the fraction of jets that are choked.
In the event that $R_*$ is larger than $R_\odot$ as considered here, we expect a larger fraction of jets to be choked, but that the resultant neutrino flux should be comparable.

Along with the distinction between successful and unsuccessful jets (see Sec.~\ref{ssec:jet properties}), the distinction between choked and visible jets described above leads to three distinct classes of jets.
These three classes are shown schematically in Fig.~\ref{fig:schematic}.

\subsection{Jet population}
We consider a population of GRBs contributing to the diffuse neutrino flux.
The population rate is described by $R(z,\Gamma_{\max})$ which we take to be a separable function for simplicity: $R(z,\Gamma_{\max})=\bar{R}(z) \xi(\Gamma_{\max})$.
The redshift dependence is taken to follow the star formation rate \cite{Yuksel:2008cu,Strolger:2015kra}: 
\begin{equation}
\bar{R}(z) \propto \left[(1+z)^{p_1 k} + \left(\frac{1+z}{5000}\right)^{p_2 k}+\left(\frac{1+z}{9}\right)^{p_3 k}\right]^{1/k}\ ,
\end{equation}
with $k=-10$, $p_1=3.4$, $p_2=-0.3$, and $p_3=-3.5$.
The $\Gamma_{\max}$ dependence is assumed to follow a power law~\cite{Denton:2017jwk}
\begin{equation}
\xi(\Gamma_{\max})=\beta_\Gamma \Gamma_{\mathrm{max}}^{\alpha_\Gamma}\ .
\end{equation}
The choice of a power law is an ansatz that satisfies the relevant criteria.
It decreases with $\Gamma$ and contains a minimal number of free parameters.
The other natural choice would be a linear fit, but a linear fit does not remain positive definite for all $\Gamma$ within the relevant range for all $\zeta_{\rm SN}$, justifying our choice of only considering a power law distribution.

Normalizing $R(z,\Gamma_{\max})$ requires two constraints~\cite{Denton:2017jwk}.
The first is the local HL-GRB rate $\bar{R}_{{\rm HL-GRB}}(z=0)=0.8$ Gpc$^{-3}$ yr$^{-1}$ \cite{Wanderman:2009es} which is the number of jets that have maximum Lorentz factors $\Gamma_{\max}\in[200,1000]$ at redshift $z=0$.
The second constraint is the local core-collapse supernova rate that forms jets, $\zeta_{\rm SN}\bar{R}_{\rm SN}(z=0)$, which is the total number of jets corrected for solid angle where $\zeta_{\rm SN}\in(0,1]$. We assume $\bar{R}_{\mathrm{SN}}(z=0) \simeq 2 \times 10^5$~Gpc$^{-3}$ yr$^{-1}$~\cite{Dahlen:2004km,Strolger:2015kra}.
These constraints are
\begin{equation}
\bar{R}_{{\rm HL-GRB}}(0)=\int_{200}^{1000}d\Gamma_{\max}\int_{\Omega(\theta<\theta_{\max})}\frac{d\Omega}{4\pi}\xi(\Gamma_{\max})\ \ \mathrm{and}\ \ \bar{R}_{\rm SN}(0)\zeta_{\rm SN}=\int_1^{1000}d\Gamma_{\max}\,\xi(\Gamma_{\max})\,.
\end{equation}
They define the GRB population's distribution and can then be used to calculate the diffuse intensity.
By comparing the intensity with observations by IceCube and ANTARES we can constrain the jet parameters $\tilde E_j$ and $\zeta_{\rm SN}$ giving a neutrino intensity compatible with the data.

\section{The low-energy excess observed by IceCube and ANTARES}\label{sec:excess}
In this Section, we introduce the low-energy events observed beyond the atmospheric background by IceCube and ANTARES.
The excess is dominant in the 30--400 TeV range and we include data down to 5 TeV as described below.
We then discuss the possibility that choked jets are main contributors to the neutrino flux observed in excess at low energies.

\subsection{The IceCube and ANTARES low-energy excess}\label{sec:excess1}
IceCube has measured high energy neutrinos from 2010 to 2016 in the High Energy Starting Event (HESE) data set \cite{Aartsen:2017mau}.
These are events where the first interaction is inside the instrumented region of the detector and not in the veto on the edges of the instrumented volume.
These events are also required to have deposited a large amount of energy in the detector.
By using these two cuts, IceCube can reduce the background rate from both muons and atmospheric neutrinos dramatically leading to a fairly pure sample.
The data set consists of two main event topologies: cascades and tracks.
The energy deposition in the cascade events is roughly spherical and is contained within the detector leading to good calorimetric energy reconstruction, while the long track events usually leave the detector providing only a lower limit on the neutrino energy.
The true energy of the initial neutrino must be then determined by an unfolding process.
Cascade events are usually connected with neutral current (NC) interactions and $\nu_e$ and $\nu_\tau$ charged current (CC) interactions while track events are usually associated with CC $\nu_\mu$ interactions.
IceCube has then reported the astrophysical flux above the background accounting for their cut efficiencies as shown in green in Fig.~\ref{fig:Inu1}.

In addition, IceCube has performed a separate analysis of the first two years of data with a lower energy threshold which has become known as the Medium Energy Starting Event (MESE) data set \cite{Aartsen:2014muf}.
With a lower threshold, two aspects of the background become problematic.
The number of background events increases considerably faster than the number of astrophysical events, and it becomes increasingly possible for a muon to pass through the veto without setting it off.
To combat these issues, the veto is dynamically extended into the detector from the outer region in a fashion that limits both types of errors occurring from low level fluctuations of individual photons.
In order to strengthen our analysis, we add in three additional low energy bins from this data set as shown in orange in Fig.~\ref{fig:Inu1}.
The higher energy bins from the MESE data set share events with the HESE data set and so we do not include them because the HESE set has more statistics.
While the 30 TeV bin dominates the data set statistically, we include the two lower energy bins for completeness; these bins have large uncertainties due to the large atmospheric backgrounds and, as such, they will constitute only a small effect on our fit.

ANTARES has also been measuring the high energy neutrino spectrum and has reported a measurement of the high energy astrophysical neutrino flux from 2007 to 2015 \cite{Albert:2017nsd}.
To reduce atmospheric backgrounds, ANTARES only considers events that are up going or within $10^\circ$ of the horizon.
The ANTARES Collaboration reported the number of events per energy bin which we then converted to diffuse intensity using the effective area of ANTARES from \cite{AdrianMartinez:2012rp} as shown in blue in Fig.~\ref{fig:Inu1}.
In order to convert the data that the ANTARES Collaboration reports to the diffuse intensity, we take the data and subtract the atmospheric backgrounds calculated by ANTARES's Monte Carlo.
It is then clear that only the 80 TeV bin is more than $1\sigma$ above the backgrounds.
We consider only the data from shower like events because the energy reconstruction is calorimetric, while for the track events ANTARES reports the energy in units determined from their Artificial Neural Network \cite{Schnabel:2013kma} which exhibits a non-trivial relationship with the true neutrino energy.

\subsection{Contribution of choked and bright jets to the neutrino excess of events}
In order to determine whether choked or bright bursts could be dominant contributors to the low-energy excess of events observed by IceCube and ANTARES, we use the latest nine year ANTARES event rate \cite{Albert:2017nsd,AdrianMartinez:2012rp} along with the low energy astrophysical signal from IceCube \cite{Aartsen:2014muf,Aartsen:2017mau}.
We then fit our advanced GRB model to the lowest energy bins from IceCube (in the $5$ TeV $<E_\nu<400$ TeV range) as well as all of the ANTARES data (in the $30$ TeV $<E_\nu<150$ TeV range) by scanning over $\tilde E_j$ and $\zeta_{\rm SN}$.
While the individual jets are normalized to the jet energy, the jet energy also affects whether or not a jet is successful, thus its feedback on the diffuse intensity is somewhat complicated.
The fraction of SNe that form jets, $\zeta_{\rm SN}$, roughly provides a total normalization to the flux.

In the following, we first compare the neutrino intensity, obtained by assuming particle acceleration at the IS radius for electromagnetically bright and choked GRBs, with the IceCube and ANTARES data. Then we discuss the possibility that the observed excess of events originates from choked jets only and include the neutrino intensity resultant from particle acceleration at the CS radius in our analysis.

\begin{figure}
\centering
\includegraphics[width=0.69\textwidth]{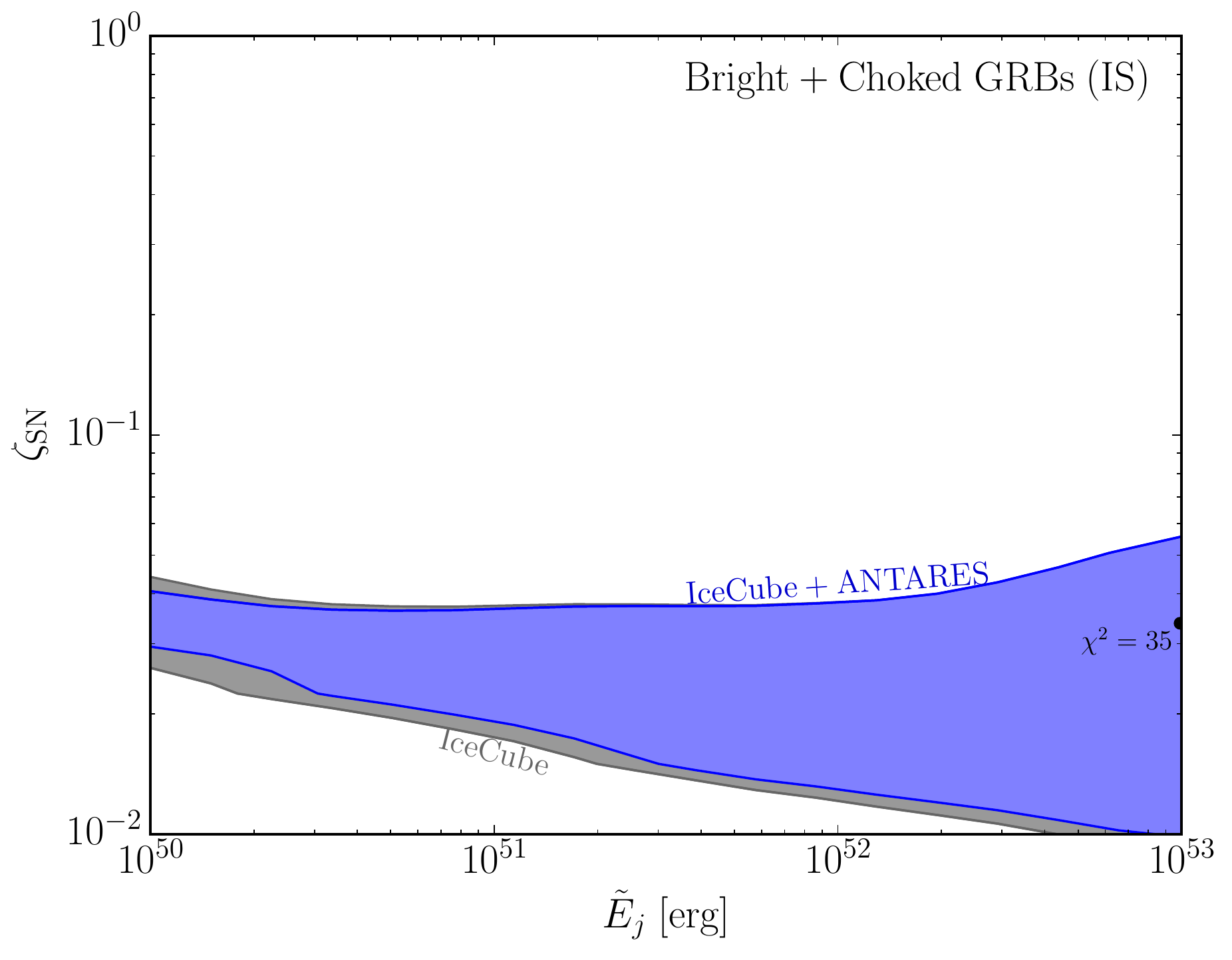}\\
\includegraphics[width=0.69\textwidth]{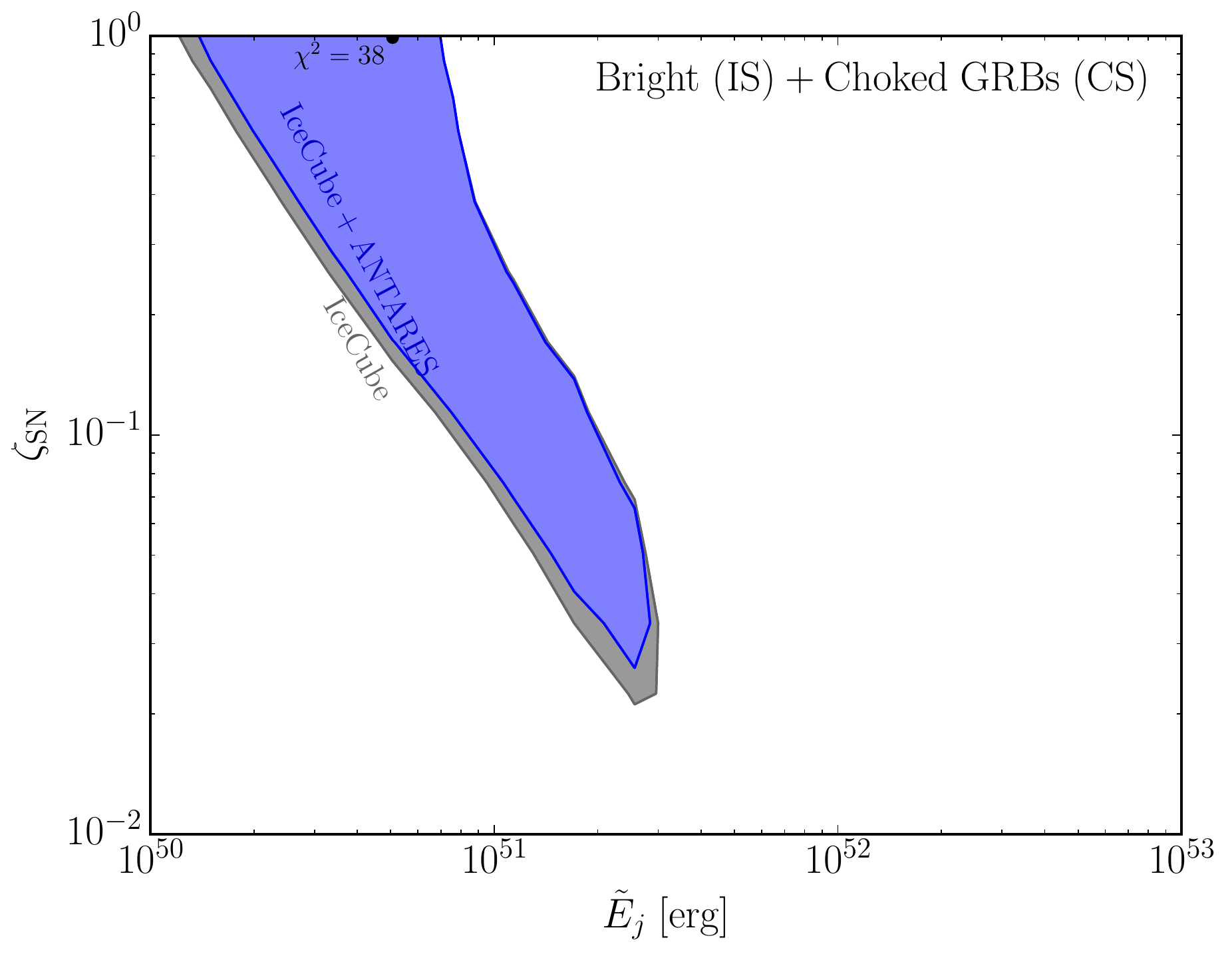}
\caption{The best fit region of parameter space in the $\zeta_{\rm SN}$ (fraction of supernovae that form jets) vs.~ $\tilde E_j$ (jet energy) space.
The shaded region is within the best fit point by $\Delta\chi^2<4.61$, the 90\% CL region for 2 d.o.f.~from the best fit point.
The cobalt blue region shows the contour for both IceCube and ANTARES data while the gray curve shows the corresponding contour for IceCube data only.
The constraint from ANTARES data alone covers most of the relevant parameter space and is not shown.
\emph{Top panel:}
Best fit region for the neutrino emission coming from choked and electromagnetically bright GRBs within the internal shock model. The best fit point has a $\chi^2_{\rm red}=3.9$; we consider it to be a poor fit.
\emph{Bottom panel:}
The same as above, but with the acceleration occurring at the collimation shock radius for choked jets and at the internal shock for visible jets.
The best fit point is at $\chi^2_{\rm red}=4.2$ which is also a poor fit.}
\label{fig:Etildej_zeta}
\end{figure}

Assuming particle acceleration at the IS radius for electromagnetically bright and choked GRBs, the top panel of Fig.~\ref{fig:Etildej_zeta} shows the $\chi^2$ for each region of the ($\tilde E_j, \zeta_{\rm SN}$) parameter space compared to the data.
The contribution to the $\chi^2$ from IceCube alone is plotted in gray along with the sum of both experiments plotted in cobalt blue.
Given the large error bars, the addition of ANTARES leads to only a small change in the best fit region.
Note that we have left the ANTARES only contour off the figure for clarity. In fact, due to poor statistics, ANTARES data alone allows for values covering nearly all of the parameter space except the lower region below $\zeta_{\rm SN}\sim0.02$ which starts to become constrained due to the bin at 70 TeV.
The best fit point is at high jet energy ($\tilde E_j = 10^{53}$~erg); the correspondent supernova fraction is $\zeta_{\rm SN}=3.4\%$.
The reduced $\chi^2$, $\chi^2_{\rm red}=\chi^2/\nu$ for $\nu=11-2$, is much larger than one $\chi^2_{\rm red}=3.9$ suggesting that it is a poor fit to the data.
The contour of the top panel of Fig.~\ref{fig:Etildej_zeta} shows the region with the smallest $\chi^2$'s, and we note that it is entirely in the region excluded by the neutrino data at higher energies as previously shown in Fig.~8 of Ref.~\cite{Denton:2017jwk}.
Moreover, we note that the majority of jets will be successful for $\tilde E_j = 10^{53}$~erg (see Fig.~\ref{fig:choked}).

\begin{figure}
\centering
\includegraphics[width=5in]{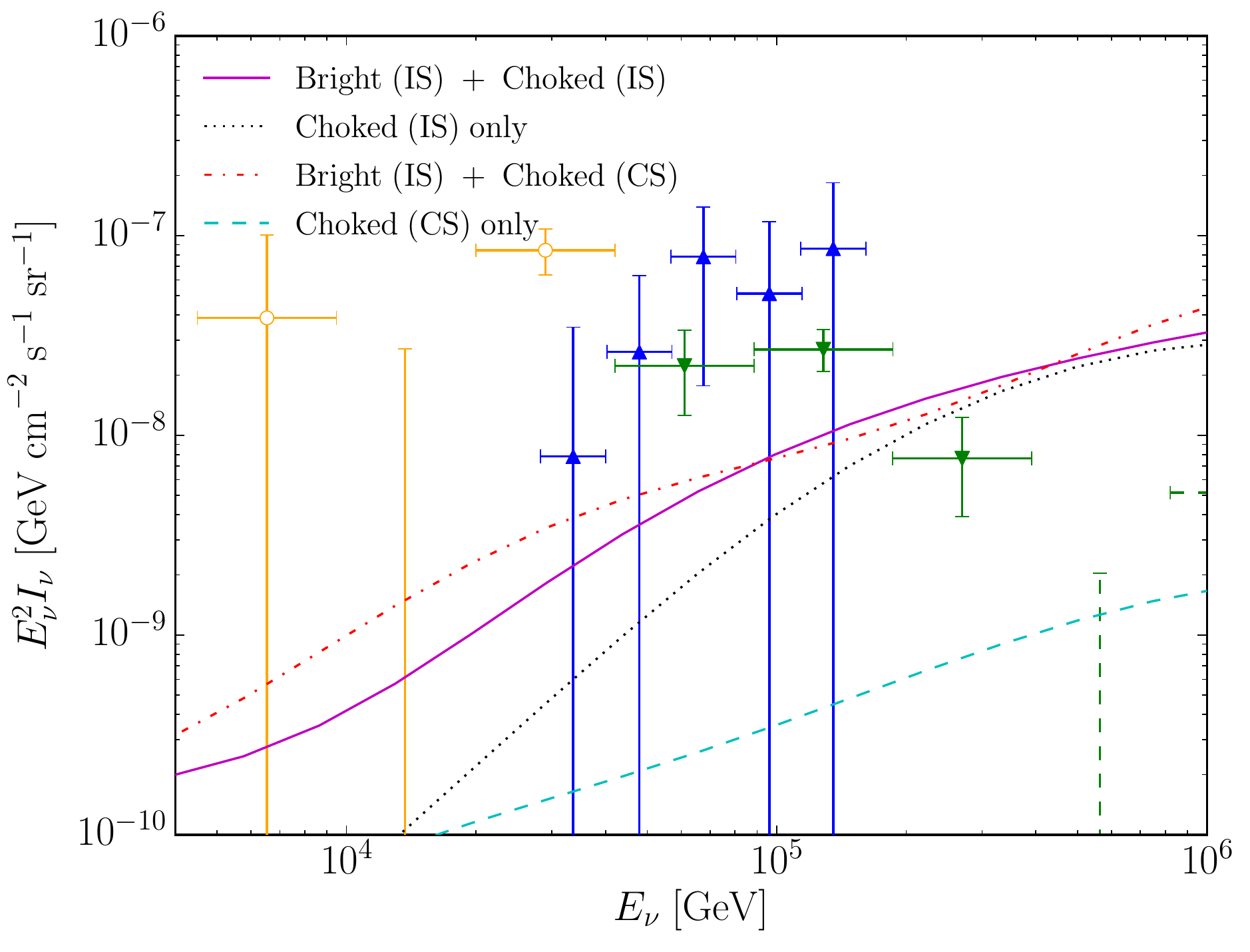}
\caption{The neutrino intensity at the best fit point from the top panel of Fig.~\ref{fig:Etildej_zeta}, ($\tilde E_j$,$\zeta_{\rm SN}$)$=$($10^{53}$~erg, $3.4\%$), is plotted in magenta.
The neutrino intensity at the IS radius from choked jets only at the best fit point from Fig.~\ref{fig:Etildej_zeta_choked}, ($\tilde E_j$,$\zeta_{\rm SN}$)$=$($1.7\times10^{51}$~erg, $58\%$), is plotted as the black dotted curve.
The neutrino intensity at the CS radius from choked jets and with (without) the visible jets at the IS is plotted in red dash-dot (cyan dashed) with best fit point ($5.1\times10^{50}$~erg, $100\%$) and ($3.8\times10^{51}$~erg, $58\%$) respectively. The data points are the same as in Fig.~\ref{fig:Inu1}, the orange circles are IceCube's MESE data \cite{Aartsen:2014muf}, the blue up triangles are the data from ANTARES \cite{Albert:2017nsd}, and the green down triangles are IceCube's HESE data \cite{Aartsen:2017mau}. 
The green dashed data points are some of the higher energy bins from IceCube's HESE data and are not included in these fits. In all investigated scenarios, the neutrino intensity describes the data poorly and it will overshoot the experimental data in the higher energy bins.}
\label{fig:Inu}
\end{figure}

Figure~\ref{fig:Inu} shows the IceCube and ANTARES data along with the neutrino intensity in magenta expected for the best fit parameters ($\tilde E_j$,$\zeta_{\rm SN}$)=($10^{53}$~erg, $3.4\%$) obtained by assuming particle acceleration at the IS radius for electromagnetically bright and choked GRBs.
One can see that the neutrino background from GRBs not only describes the neutrino data poorly, it will clearly overshoot the higher energy bins which are not included in this analysis; see Fig.~5 of Ref.~\cite{Denton:2017jwk} for a plot of the expected neutrino intensity at higher neutrino energies.

\begin{figure}
\centering
\includegraphics[width=0.69\textwidth]{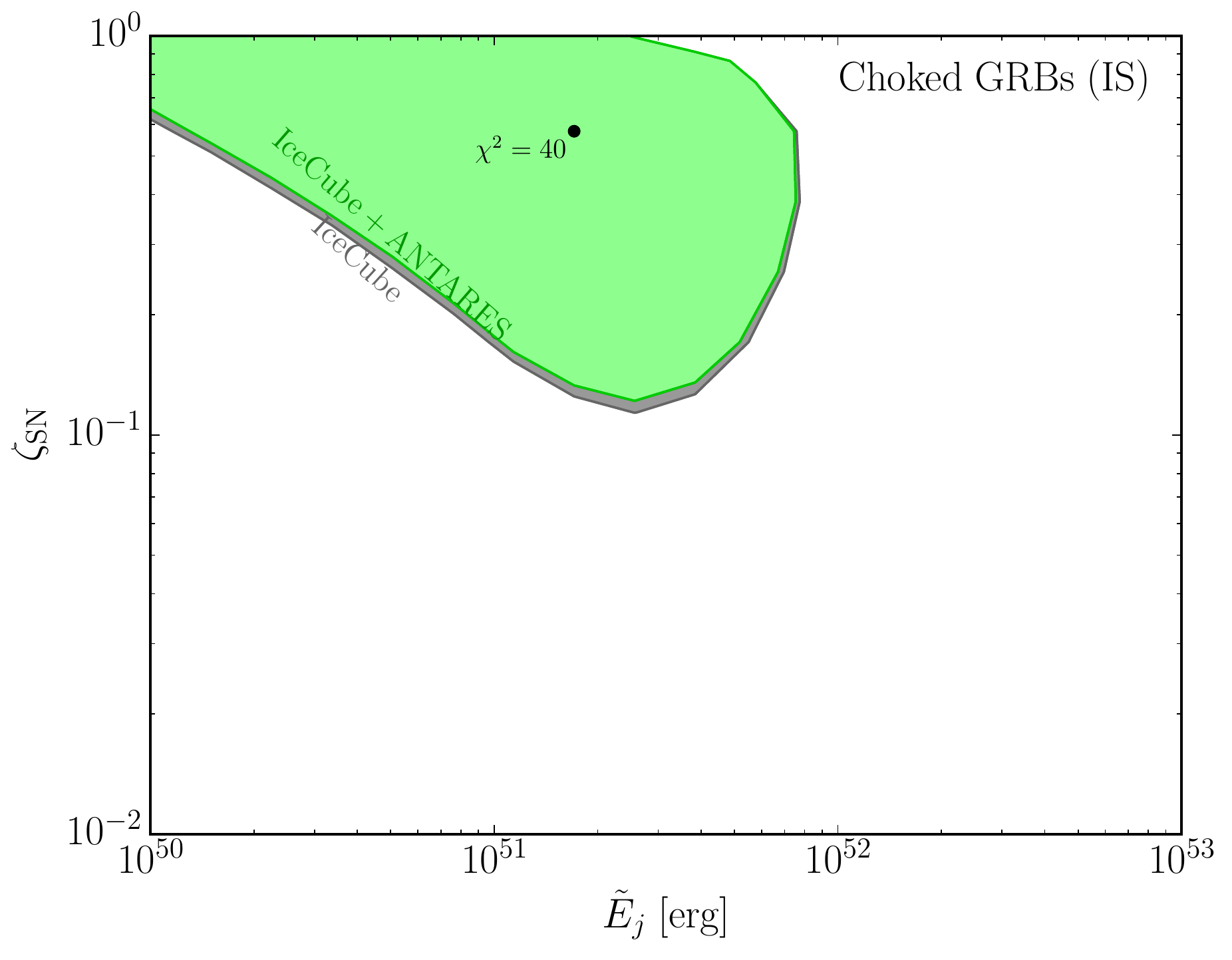}
\caption{The same as Fig.~\ref{fig:Etildej_zeta} except based on the flux from choked jets only.
Particle acceleration is assumed to happen at the IS radius. The best fit point has a $\chi^2_{\rm red}=4.4$; we consider it to be a poor fit.
When particle acceleration is assumed to happen at the CS radius, the $\chi^2$ is very shallow and no particular region can be constrained; the minimum $\chi^2_{\rm red}=5.4$, which is a worse fit than the the internal shock model.}
\label{fig:Etildej_zeta_choked}
\end{figure}

As discussed in Sec.~\ref{sec:introduction}, choked sources have been suggested as possible candidates to interpret the observed low-energy excess.
To this purpose, as further check, in Fig.~\ref{fig:Etildej_zeta_choked} we show the best fit region of parameter space in the ($\zeta_{\rm SN}$, $\tilde E_j$) similarly to Fig.~\ref{fig:Etildej_zeta} but for choked jets only.
We consider particle acceleration at the IS radius; the best fit point has now changed with respect to the top panel of Fig.~\ref{fig:Etildej_zeta}, but the data are still poorly fit ($\chi^2_{\rm red}=4.4$).
Correspondingly, Fig.~\ref{fig:Inu} shows the neutrino intensity from choked jets only in dashed black for the best fit parameters for choked jets ($\tilde E_j$,$\zeta_{\rm SN}$)=($1.7\e{51}$~erg, $58\%$).
Also from here, it is clear that choked GRBs do not describe the data well.

We also calculated the $\chi^2$ for choked jets only where the acceleration occurs at the CS radius, but since the $\chi^2$ is so shallow, all regions of parameter space are allowed at roughly the same significance (plot not shown). 
The minimum $\chi^2_{\rm red}$ is $5.4$ at ($\tilde E_j$,$\zeta_{\rm SN}$)=($3.8\e{51}$~erg, $100\%$) in this case. The corresponding neutrino intensity for the best-fit ($\tilde E_j$,$\zeta_{\rm SN}$) is shown in Fig.~\ref{fig:Inu} in cyan dashed.

Observational constraints would not be violated if particle acceleration should occur at the IS radius for electromagnetically bright sources and at the CS radius for choked bursts. In order to make our conclusions more robust, the bottom panel of Fig.~\ref{fig:Etildej_zeta} indeed shows the best fit region for acceleration occurring at the CS radius for choked jets and at the IS radius for electromagnetically bright jets. 
The minimum $\chi^2_{\rm red}$ is $4.2$ in this case for ($\tilde E_j$,$\zeta_{\rm SN}$)=($5.1\e{50}$~erg, $100\%$) and the correspondent neutrino intensity is shown in red dash-dot in Fig.~\ref{fig:Inu}. 
One can see that also this scenario fits the neutrino data poorly.

Note that the neutrino data sample considered in this paper is different from the one adopted in Ref.~\cite{Denton:2017jwk}.
The addition of the MESE data does not significantly modify the shape of the contours in Figs.~\ref{fig:Etildej_zeta} and \ref{fig:Etildej_zeta_choked}, but it does worsen the quality of fit significantly.
In particular, the 30 TeV data point from the MESE sample is rather high which is very difficult to fit even with neutrinos from choked GRBs.

\section{Conclusions and outlook}\label{sec:conclusions}
There is mounting evidence that the astrophysical neutrino flux observed in the TeV-PeV energy range is not due to one class of sources. An apparent excess of events over the atmospheric background at lower energies (30--400 TeV) is observed by both the IceCube and ANTARES Telescopes~\cite{Aartsen:2014muf,Aartsen:2015zva,Albert:2017nsd,Aartsen:2017mau} suggesting that a broken power law may be a better fit to the observed data.
Hidden astrophysical sources, i.e.~sources not electromagnetically bright, have been put forward to explain the low-energy component of the observed neutrino spectrum and alleviate tension with {\it Fermi} data~\cite{Murase:2015xka}.

In Ref.~\cite{Denton:2017jwk}, we presented an advanced modeling of the neutrino emission from electromagnetically bright and choked gamma-ray bursts and found that current data suggest the majority of the jets to be choked. In the light of these findings, in this paper, we investigate whether neutrinos from bright and choked gamma-ray bursts can describe well the low-energy excess of events observed by IceCube and ANTARES. Since the model proposed in Ref.~\cite{Denton:2017jwk} is very general by construction, it represents an optimal framework to test whether astrophysical bursts can explain the observed low-energy neutrino excess without the addition new ad-hoc parameters. 

If particle acceleration at the collimation shock is efficient, the resultant neutrino intensity may peak at lower energies than the one coming from particles accelerated at the internal shock. Since in this work we focus on the low-energy excess of neutrino events, we also investigate the particle production and acceleration at the collimation shock radius beyond the one at the internal shock radius. 
We find that the collimation shock radius is larger than the internal shock radius for most of the scanned parameter space.
Hence, particle acceleration cannot happen at the collimation shock radius as it would violate the observed variability times of electromagnetically bright GRBs.
Assuming that particle acceleration occurs with full efficiency at the larger radius, we therefore consider that all the particle production and acceleration can be confined to the internal shock radius in the advanced gamma-ray burst model.

Since there is no variability time information for choked jets, we have also investigated the effect of particle acceleration at the collimation shock radius for choked jets only, while particles in visible jets are still accelerated at the internal shock radius.
However, acceleration at the collimation shock radius for choked GRBs slightly worsens the fit to the data compared to that of the internal shock radius.

Due to the fact that the neutrino diffuse background from astrophysical jets predicted within the advanced gamma-ray burst model peaks at higher energies, the predicted neutrino spectrum is much harder than the rather soft spectrum observed by IceCube and ANTARES, especially from the MESE data set.
In fact, even our best fit neutrino intensity would subsequently overshoot the higher energy bins (not included in the fit in this paper).
This is largely due to the fact that we have extended the Amati relation down to choked GRBs.
While the Amati relation was originally derived for high-luminosity GRBs, we assume that it naturally extrapolates to choked GRBs.
The only alternative would be to fit the GRB properties to the IceCube and ANTARES data, at which point any neutrino flux could be obtained.
We choose instead to keep our model connected with GRB observations as much as possible.

Our findings imply that a fine tuning of the model parameters or distinct scenarios for hidden sources should be invoked in order to explain the measured low-energy neutrino flux. 
We also compared a fit of all neutrinos produced in our model from bright and choked jets to that of neutrinos produced only in choked jets at the internal and collimation shock radii, finding that the choked-only case fits the data even worse than in the case where all neutrinos were included.
As pointed out in Ref.~\cite{Denton:2017jwk} most of the neutrino production, even in the case of choked jets, it is driven by $p$--$\gamma$ interactions in the advanced gamma-ray burst model.
If a sizable amount of neutrinos should be produced through $p$--$p$ interactions in the jet, this may be responsible for a neutrino intensity at lower energies larger than the one foreseen within the advanced gamma-ray burst model. Mechanisms enhancing the $p$--$p$ efficiency within choked bursts or calorimeters such as starburst galaxies should be further explored in order to unveil the origin of the low-energy excess observed in the data.

As IceCube and ANTARES collect more data, and with the advent of future detectors such as KM3NeT \cite{Adrian-Martinez:2016fdl} and IceCube-Gen2 \cite{Ackermann:2017pja} increasingly robust constraints could be placed on the theoretical models estimating the neutrino production in astrophysical sources.
In fact precisely determining both the normalization and the shape of the high-energy neutrino flux will be vital for understanding the origins of high energy neutrinos. In turn such information will also allow us to better constrain the physics of the sources.

\acknowledgments
We thank Markus Ahlers for helpful comments on the manuscript.
PBD and IT acknowledge support from the Villum Foundation (Project No.~13164), and by the Danish National Research Foundation (DNRF91).
PBD thanks the Danish National Research Foundation (Grant No.~1041811001) for support. The work of IT has also been supported by the Knud H\o jgaard Foundation and the Deutsche Forschungsgemeinschaft through Sonderforschungbereich SFB 1258 ``Neutrinos and Dark Matter in Astro- and Particle Physics'' (NDM). 

\bibliographystyle{JHEP}
\bibliography{GRB}

\providecommand{\href}[2]{#2}\begingroup\raggedright\begin{thebibliography}{10}

\bibitem{Aartsen:2013bka}
{\scshape IceCube} collaboration, M.~G. Aartsen et~al., \emph{{First
  observation of PeV-energy neutrinos with IceCube}},
  \href{http://dx.doi.org/10.1103/PhysRevLett.111.021103}{\emph{Phys. Rev.
  Lett.} {\bf 111} (2013) 021103}, [\href{https://arxiv.org/abs/1304.5356}{{\tt
  1304.5356}}].

\bibitem{Aartsen:2013jdh}
{\scshape IceCube} collaboration, M.~G. Aartsen et~al., \emph{{Evidence for
  High-Energy Extraterrestrial Neutrinos at the IceCube Detector}},
  \href{http://dx.doi.org/10.1126/science.1242856}{\emph{Science} {\bf 342}
  (2013) 1242856}, [\href{https://arxiv.org/abs/1311.5238}{{\tt 1311.5238}}].

\bibitem{Aartsen:2014gkd}
{\scshape IceCube} collaboration, M.~G. Aartsen et~al., \emph{{Observation of
  High-Energy Astrophysical Neutrinos in Three Years of IceCube Data}},
  \href{http://dx.doi.org/10.1103/PhysRevLett.113.101101}{\emph{Phys. Rev.
  Lett.} {\bf 113} (2014) 101101}, [\href{https://arxiv.org/abs/1405.5303}{{\tt
  1405.5303}}].

\bibitem{Aartsen:2014muf}
{\scshape IceCube} collaboration, M.~G. Aartsen et~al., \emph{{Atmospheric and
  astrophysical neutrinos above 1 TeV interacting in IceCube}},
  \href{http://dx.doi.org/10.1103/PhysRevD.91.022001}{\emph{Phys. Rev.} {\bf
  D91} (2015) 022001}, [\href{https://arxiv.org/abs/1410.1749}{{\tt
  1410.1749}}].

\bibitem{Aartsen:2015knd}
{\scshape IceCube} collaboration, M.~G. Aartsen et~al., \emph{{A combined
  maximum-likelihood analysis of the high-energy astrophysical neutrino flux
  measured with IceCube}},
  \href{http://dx.doi.org/10.1088/0004-637X/809/1/98}{\emph{Astrophys. J.} {\bf
  809} (2015) 98}, [\href{https://arxiv.org/abs/1507.03991}{{\tt 1507.03991}}].

\bibitem{Aartsen:2015rwa}
{\scshape IceCube} collaboration, M.~G. Aartsen et~al., \emph{{Evidence for
  Astrophysical Muon Neutrinos from the Northern Sky with IceCube}},
  \href{http://dx.doi.org/10.1103/PhysRevLett.115.081102}{\emph{Phys. Rev.
  Lett.} {\bf 115} (2015) 081102},
  [\href{https://arxiv.org/abs/1507.04005}{{\tt 1507.04005}}].

\bibitem{Aartsen:2015zva}
{\scshape IceCube} collaboration, M.~G. Aartsen et~al., \emph{{The IceCube
  Neutrino Observatory - Contributions to ICRC 2015 Part II: Atmospheric and
  Astrophysical Diffuse Neutrino Searches of All Flavors}},  in
  \emph{{Proceedings, 34th International Cosmic Ray Conference (ICRC 2015): The
  Hague, The Netherlands, July 30-August 6, 2015}}, 2015.
\newblock \href{https://arxiv.org/abs/1510.05223}{{\tt 1510.05223}}.

\bibitem{Aartsen:2017mau}
{\scshape IceCube} collaboration, M.~G. Aartsen et~al., \emph{{The IceCube
  Neutrino Observatory - Contributions to ICRC 2017 Part II: Properties of the
  Atmospheric and Astrophysical Neutrino Flux}},
  \href{https://arxiv.org/abs/1710.01191}{{\tt 1710.01191}}.

\bibitem{Palladino:2016zoe}
A.~Palladino and F.~Vissani, \emph{{Extragalactic plus Galactic model for
  IceCube neutrino events}},
  \href{http://dx.doi.org/10.3847/0004-637X/826/2/185}{\emph{Astrophys. J.}
  {\bf 826} (2016) 185}, [\href{https://arxiv.org/abs/1601.06678}{{\tt
  1601.06678}}].

\bibitem{Denton:2017csz}
P.~B. Denton, D.~Marfatia and T.~J. Weiler, \emph{{The Galactic Contribution to
  IceCube's Astrophysical Neutrino Flux}},
  \href{http://dx.doi.org/10.1088/1475-7516/2017/08/033}{\emph{JCAP} {\bf 1708}
  (2017) 033}, [\href{https://arxiv.org/abs/1703.09721}{{\tt 1703.09721}}].

\bibitem{Aartsen:2017ujz}
{\scshape IceCube} collaboration, M.~G. Aartsen et~al., \emph{{Constraints on
  Galactic Neutrino Emission with Seven Years of IceCube Data}},
  \href{http://dx.doi.org/10.3847/1538-4357/aa8dfb}{\emph{Astrophys. J.} {\bf
  849} (2017) 67}, [\href{https://arxiv.org/abs/1707.03416}{{\tt 1707.03416}}].

\bibitem{Meszaros:2015krr}
P.~M\'esz\'aros, \emph{{Gamma Ray Bursts as Neutrino Sources}},  pp.~1--14.
\newblock 2017.
\newblock \href{https://arxiv.org/abs/1511.01396}{{\tt 1511.01396}}.
\newblock \href{http://dx.doi.org/10.1142/9789814759410_0001}{DOI}.

\bibitem{Waxman:2015ues}
E.~Waxman, \emph{{The Origin of IceCube's Neutrinos: Cosmic Ray Accelerators
  Embedded in Star Forming Calorimeters}},  pp.~33--45.
\newblock 2017.
\newblock \href{https://arxiv.org/abs/1511.00815}{{\tt 1511.00815}}.
\newblock \href{http://dx.doi.org/10.1142/9789814759410_0003}{DOI}.

\bibitem{Murase:2015ndr}
K.~Murase, \emph{{Active Galactic Nuclei as High-Energy Neutrino Sources}},
  pp.~15--31.
\newblock 2017.
\newblock \href{https://arxiv.org/abs/1511.01590}{{\tt 1511.01590}}.
\newblock \href{http://dx.doi.org/10.1142/9789814759410_0002}{DOI}.

\bibitem{Ahlers:2015lln}
M.~Ahlers and F.~Halzen, \emph{{High-energy cosmic neutrino puzzle: a review}},
  \href{http://dx.doi.org/10.1088/0034-4885/78/12/126901}{\emph{Rept. Prog.
  Phys.} {\bf 78} (2015) 126901}.

\bibitem{Anchordoqui:2013dnh}
L.~A. Anchordoqui et~al., \emph{{Cosmic Neutrino Pevatrons: A Brand New Pathway
  to Astronomy, Astrophysics, and Particle Physics}},
  \href{http://dx.doi.org/10.1016/j.jheap.2014.01.001}{\emph{JHEAp} {\bf 1-2}
  (2014) 1--30}, [\href{https://arxiv.org/abs/1312.6587}{{\tt 1312.6587}}].

\bibitem{Dai:2016gtz}
L.~Dai and K.~Fang, \emph{{Can tidal disruption events produce the IceCube
  neutrinos?}}, \href{http://dx.doi.org/10.1093/mnras/stx863}{\emph{Mon. Not.
  Roy. Astron. Soc.} {\bf 469} (2017) 1354--1359},
  [\href{https://arxiv.org/abs/1612.00011}{{\tt 1612.00011}}].

\bibitem{Senno:2016bso}
N.~Senno, K.~Murase and P.~Meszaros, \emph{{High-energy Neutrino Flares from
  X-Ray Bright and Dark Tidal Disruption Events}},
  \href{http://dx.doi.org/10.3847/1538-4357/aa6344}{\emph{Astrophys. J.} {\bf
  838} (2017) 3}, [\href{https://arxiv.org/abs/1612.00918}{{\tt 1612.00918}}].

\bibitem{Lunardini:2016xwi}
C.~Lunardini and W.~Winter, \emph{{High Energy Neutrinos from the Tidal
  Disruption of Stars}},
  \href{http://dx.doi.org/10.1103/PhysRevD.95.123001}{\emph{Phys. Rev.} {\bf
  D95} (2017) 123001}, [\href{https://arxiv.org/abs/1612.03160}{{\tt
  1612.03160}}].

\bibitem{Chen:2014gxa}
C.-Y. Chen, P.~S. Bhupal~Dev and A.~Soni, \emph{{Two-component flux explanation
  for the high energy neutrino events at IceCube}},
  \href{http://dx.doi.org/10.1103/PhysRevD.92.073001}{\emph{Phys. Rev.} {\bf
  D92} (2015) 073001}, [\href{https://arxiv.org/abs/1411.5658}{{\tt
  1411.5658}}].

\bibitem{Chianese:2016opp}
M.~Chianese, G.~Miele, S.~Morisi and E.~Vitagliano, \emph{{Low energy IceCube
  data and a possible Dark Matter related excess}},
  \href{http://dx.doi.org/10.1016/j.physletb.2016.03.084}{\emph{Phys. Lett.}
  {\bf B757} (2016) 251--256}, [\href{https://arxiv.org/abs/1601.02934}{{\tt
  1601.02934}}].

\bibitem{Anchordoqui:2016ewn}
L.~A. Anchordoqui, M.~M. Block, L.~Durand, P.~Ha, J.~F. Soriano and T.~J.
  Weiler, \emph{{Evidence for a break in the spectrum of astrophysical
  neutrinos}}, \href{http://dx.doi.org/10.1103/PhysRevD.95.083009}{\emph{Phys.
  Rev.} {\bf D95} (2017) 083009}, [\href{https://arxiv.org/abs/1611.07905}{{\tt
  1611.07905}}].

\bibitem{Palladino:2017qda}
A.~Palladino, C.~Mascaretti and F.~Vissani, \emph{{On the compatibility of the
  IceCube results with a universal neutrino spectrum}},
  \href{http://dx.doi.org/10.1140/epjc/s10052-017-5273-z}{\emph{Eur. Phys. J.}
  {\bf C77} (2017) 684}, [\href{https://arxiv.org/abs/1708.02094}{{\tt
  1708.02094}}].

\bibitem{Palladino:2018evm}
A.~Palladino and W.~Winter, \emph{{A Multi-Component Model for the Observed
  Astrophysical Neutrinos}},  \href{https://arxiv.org/abs/1801.07277}{{\tt
  1801.07277}}.

\bibitem{Albert:2017nsd}
{\scshape ANTARES} collaboration, A.~Albert et~al., \emph{{All-flavor Search
  for a Diffuse Flux of Cosmic Neutrinos with Nine Years of ANTARES Data}},
  \href{http://dx.doi.org/10.3847/2041-8213/aaa4f6}{\emph{Astrophys. J.} {\bf
  853} (2018) L7}, [\href{https://arxiv.org/abs/1711.07212}{{\tt 1711.07212}}].

\bibitem{Chianese:2017jfa}
M.~Chianese, R.~Mele, G.~Miele, P.~Migliozzi and S.~Morisi, \emph{{Use of
  ANTARES and IceCube Data to Constrain a Single Power-law Neutrino Flux}},
  \href{http://dx.doi.org/10.3847/1538-4357/aa97e6}{\emph{Astrophys. J.} {\bf
  851} (2017) 36}, [\href{https://arxiv.org/abs/1707.05168}{{\tt 1707.05168}}].

\bibitem{AdrianMartinez:2012rp}
{\scshape ANTARES} collaboration, S.~Adrian-Martinez et~al., \emph{{Search for
  Cosmic Neutrino Point Sources with Four Year Data of the ANTARES Telescope}},
  \href{http://dx.doi.org/10.1088/0004-637X/760/1/53}{\emph{Astrophys. J.} {\bf
  760} (2012) 53}, [\href{https://arxiv.org/abs/1207.3105}{{\tt 1207.3105}}].

\bibitem{Murase:2015xka}
K.~Murase, D.~Guetta and M.~Ahlers, \emph{{Hidden Cosmic-Ray Accelerators as an
  Origin of TeV-PeV Cosmic Neutrinos}},
  \href{http://dx.doi.org/10.1103/PhysRevLett.116.071101}{\emph{Phys. Rev.
  Lett.} {\bf 116} (2016) 071101},
  [\href{https://arxiv.org/abs/1509.00805}{{\tt 1509.00805}}].

\bibitem{Ahlers:2014ioa}
M.~Ahlers and F.~Halzen, \emph{{Pinpointing Extragalactic Neutrino Sources in
  Light of Recent IceCube Observations}},
  \href{http://dx.doi.org/10.1103/PhysRevD.90.043005}{\emph{Phys. Rev.} {\bf
  D90} (2014) 043005}, [\href{https://arxiv.org/abs/1406.2160}{{\tt
  1406.2160}}].

\bibitem{Ando:2015bva}
S.~Ando, I.~Tamborra and F.~Zandanel, \emph{{Tomographic Constraints on
  High-Energy Neutrinos of Hadronuclear Origin}},
  \href{http://dx.doi.org/10.1103/PhysRevLett.115.221101}{\emph{Phys. Rev.
  Lett.} {\bf 115} (2015) 221101},
  [\href{https://arxiv.org/abs/1509.02444}{{\tt 1509.02444}}].

\bibitem{Bechtol:2015uqb}
K.~Bechtol, M.~Ahlers, M.~Di~Mauro, M.~Ajello and J.~Vandenbroucke,
  \emph{{Evidence against star-forming galaxies as the dominant source of
  IceCube neutrinos}},
  \href{http://dx.doi.org/10.3847/1538-4357/836/1/47}{\emph{Astrophys. J.} {\bf
  836} (2017) 47}, [\href{https://arxiv.org/abs/1511.00688}{{\tt 1511.00688}}].

\bibitem{Murase:2016gly}
K.~Murase and E.~Waxman, \emph{{Constraining High-Energy Cosmic Neutrino
  Sources: Implications and Prospects}},
  \href{http://dx.doi.org/10.1103/PhysRevD.94.103006}{\emph{Phys. Rev.} {\bf
  D94} (2016) 103006}, [\href{https://arxiv.org/abs/1607.01601}{{\tt
  1607.01601}}].

\bibitem{Mertsch:2016hcd}
P.~Mertsch, M.~Rameez and I.~Tamborra, \emph{{Detection prospects for high
  energy neutrino sources from the anisotropic matter distribution in the local
  universe}},
  \href{http://dx.doi.org/10.1088/1475-7516/2017/03/011}{\emph{JCAP} {\bf 1703}
  (2017) 011}, [\href{https://arxiv.org/abs/1612.07311}{{\tt 1612.07311}}].

\bibitem{Meszaros:2001ms}
P.~M\'esz\'aros and E.~Waxman, \emph{{TeV neutrinos from successful and choked
  gamma-ray bursts}},
  \href{http://dx.doi.org/10.1103/PhysRevLett.87.171102}{\emph{Phys. Rev.
  Lett.} {\bf 87} (2001) 171102},
  [\href{https://arxiv.org/abs/astro-ph/0103275}{{\tt astro-ph/0103275}}].

\bibitem{Ando:2005xi}
S.~Ando and J.~F. Beacom, \emph{{Revealing the supernova-gamma-ray burst
  connection with TeV neutrinos}},
  \href{http://dx.doi.org/10.1103/PhysRevLett.95.061103}{\emph{Phys. Rev.
  Lett.} {\bf 95} (2005) 061103},
  [\href{https://arxiv.org/abs/astro-ph/0502521}{{\tt astro-ph/0502521}}].

\bibitem{Horiuchi:2007xi}
S.~Horiuchi and S.~Ando, \emph{{High-energy neutrinos from reverse shocks in
  choked and successful relativistic jets}},
  \href{http://dx.doi.org/10.1103/PhysRevD.77.063007}{\emph{Phys. Rev.} {\bf
  D77} (2008) 063007}, [\href{https://arxiv.org/abs/0711.2580}{{\tt
  0711.2580}}].

\bibitem{Razzaque:2003uw}
S.~Razzaque, P.~M\'esz\'aros and E.~Waxman, \emph{{Neutrino signatures of the
  supernova - gamma-ray burst relationship}},
  \href{http://dx.doi.org/10.1103/PhysRevD.69.023001}{\emph{Phys. Rev.} {\bf
  D69} (2004) 023001}, [\href{https://arxiv.org/abs/astro-ph/0308239}{{\tt
  astro-ph/0308239}}].

\bibitem{Razzaque:2003uv}
S.~Razzaque, P.~M\'esz\'aros and E.~Waxman, \emph{{Neutrino tomography of
  gamma-ray bursts and massive stellar collapses}},
  \href{http://dx.doi.org/10.1103/PhysRevD.68.083001}{\emph{Phys. Rev.} {\bf
  D68} (2003) 083001}, [\href{https://arxiv.org/abs/astro-ph/0303505}{{\tt
  astro-ph/0303505}}].

\bibitem{Razzaque:2002kb}
S.~Razzaque, P.~M\'esz\'aros and E.~Waxman, \emph{{High energy neutrinos from
  gamma-ray bursts with precursor supernovae}},
  \href{http://dx.doi.org/10.1103/PhysRevLett.90.241103}{\emph{Phys. Rev.
  Lett.} {\bf 90} (2003) 241103},
  [\href{https://arxiv.org/abs/astro-ph/0212536}{{\tt astro-ph/0212536}}].

\bibitem{Tamborra:2015fzv}
I.~Tamborra and S.~Ando, \emph{{Inspecting the supernova–gamma-ray-burst
  connection with high-energy neutrinos}},
  \href{http://dx.doi.org/10.1103/PhysRevD.93.053010}{\emph{Phys. Rev.} {\bf
  D93} (2016) 053010}, [\href{https://arxiv.org/abs/1512.01559}{{\tt
  1512.01559}}].

\bibitem{Senno:2015tsn}
N.~Senno, K.~Murase and P.~M\'esz\'aros, \emph{{Choked Jets and Low-Luminosity
  Gamma-Ray Bursts as Hidden Neutrino Sources}},
  \href{http://dx.doi.org/10.1103/PhysRevD.93.083003}{\emph{Phys. Rev.} {\bf
  D93} (2016) 083003}, [\href{https://arxiv.org/abs/1512.08513}{{\tt
  1512.08513}}].

\bibitem{Senno:2017vtd}
N.~Senno, K.~Murase and P.~M\'esz\'aros, \emph{{Constraining high-energy
  neutrino emission from choked jets in stripped-envelope supernovae}},
  \href{http://dx.doi.org/10.1088/1475-7516/2018/01/025}{\emph{JCAP} {\bf 1801}
  (2018) 025}, [\href{https://arxiv.org/abs/1706.02175}{{\tt 1706.02175}}].

\bibitem{Murase:2013ffa}
K.~Murase and K.~Ioka, \emph{{TeV–PeV Neutrinos from Low-Power Gamma-Ray
  Burst Jets inside Stars}},
  \href{http://dx.doi.org/10.1103/PhysRevLett.111.121102}{\emph{Phys. Rev.
  Lett.} {\bf 111} (2013) 121102}, [\href{https://arxiv.org/abs/1306.2274}{{\tt
  1306.2274}}].

\bibitem{Denton:2017jwk}
P.~B. Denton and I.~Tamborra, \emph{{Exploring the Properties of Choked
  Gamma-Ray Bursts with IceCube's High Energy Neutrinos}},
  \href{http://dx.doi.org/10.3847/1538-4357/aaab4a}{\emph{Astrophys. J.} {\bf
  855} (2018) 37}, [\href{https://arxiv.org/abs/1711.00470}{{\tt 1711.00470}}].

\bibitem{Biehl:2017qen}
D.~Biehl, J.~Heinze and W.~Winter, \emph{{Expected neutrino fluence from short
  Gamma-Ray Burst 170817A and off-axis angle constraints}},
  \href{http://dx.doi.org/10.1093/mnras/sty285}{\emph{Mon. Not. Roy. Astron.
  Soc.} (2018) }, [\href{https://arxiv.org/abs/1712.00449}{{\tt 1712.00449}}].

\bibitem{Gehrels:2013xd}
N.~Gehrels and S.~Razzaque, \emph{{Gamma Ray Bursts in the Swift-Fermi Era}},
  \href{http://dx.doi.org/10.1007/s11467-013-0282-3}{\emph{Front.
  Phys.(Beijing)} {\bf 8} (2013) 661--678},
  [\href{https://arxiv.org/abs/1301.0840}{{\tt 1301.0840}}].

\bibitem{Lu:2017toj}
H.~Lü, X.~Wang, R.~Lu, L.~Lan, H.~Gao, E.~Liang et~al., \emph{{A Peculiar GRB
  110731A: Lorentz Factor, Jet Composition, Central Engine, and Progenitor}},
  \href{http://dx.doi.org/10.3847/1538-4357/aa78f0}{\emph{Astrophys. J.} {\bf
  843} (2017) 114}, [\href{https://arxiv.org/abs/1706.00898}{{\tt
  1706.00898}}].

\bibitem{Band:1993eg}
D.~Band et~al., \emph{{BATSE observations of gamma-ray burst spectra. 1.
  Spectral diversity.}},
  \href{http://dx.doi.org/10.1086/172995}{\emph{Astrophys. J.} {\bf 413} (1993)
  281--292}.

\bibitem{Amati:2002ny}
L.~Amati et~al., \emph{{Intrinsic spectra and energetics of BeppoSAX gamma-ray
  bursts with known redshifts}},
  \href{http://dx.doi.org/10.1051/0004-6361:20020722}{\emph{Astron. Astrophys.}
  {\bf 390} (2002) 81}, [\href{https://arxiv.org/abs/astro-ph/0205230}{{\tt
  astro-ph/0205230}}].

\bibitem{Yonetoku:2003gi}
D.~Yonetoku, T.~Murakami, T.~Nakamura, R.~Yamazaki, A.~K. Inoue and K.~Ioka,
  \emph{{Gamma-ray burst formation rates inferred from the spectral peak
  energy-peak luminosity relation}},
  \href{http://dx.doi.org/10.1086/421285}{\emph{Astrophys. J.} {\bf 609} (2004)
  935}, [\href{https://arxiv.org/abs/astro-ph/0309217}{{\tt
  astro-ph/0309217}}].

\bibitem{Olive:2016xmw}
{\scshape Particle Data Group} collaboration, C.~Patrignani et~al.,
  \emph{{Review of Particle Physics}},
  \href{http://dx.doi.org/10.1088/1674-1137/40/10/100001}{\emph{Chin. Phys.}
  {\bf C40} (2016) 100001}.

\bibitem{Bromberg:2011fg}
O.~Bromberg, E.~Nakar, T.~Piran and R.~Sari, \emph{{The propagation of
  relativistic jets in external media}},
  \href{http://dx.doi.org/10.1088/0004-637X/740/2/100}{\emph{Astrophys. J.}
  {\bf 740} (2011) 100}, [\href{https://arxiv.org/abs/1107.1326}{{\tt
  1107.1326}}].

\bibitem{Thompson:1994zh}
C.~Thompson, \emph{{A Model of gamma-ray bursts}}, {\emph{Mon. Not. Roy.
  Astron. Soc.} {\bf 270} (1994) 480}.

\bibitem{Biehl:2017zlw}
D.~Biehl, D.~Boncioli, A.~Fedynitch and W.~Winter, \emph{{Cosmic-Ray and
  Neutrino Emission from Gamma-Ray Bursts with a Nuclear Cascade}},
  \href{https://arxiv.org/abs/1705.08909}{{\tt 1705.08909}}.

\bibitem{Zhang:2010jt}
B.~Zhang and H.~Yan, \emph{{The Internal-Collision-Induced Magnetic
  Reconnection and Turbulence (ICMART) Model of Gamma-Ray Bursts}},
  \href{http://dx.doi.org/10.1088/0004-637X/726/2/90}{\emph{Astrophys. J.} {\bf
  726} (2011) 90}, [\href{https://arxiv.org/abs/1011.1197}{{\tt 1011.1197}}].

\bibitem{Zhang:2012qy}
B.~Zhang and P.~Kumar, \emph{{Model-dependent high-energy neutrino flux from
  Gamma-Ray Bursts}},
  \href{http://dx.doi.org/10.1103/PhysRevLett.110.121101}{\emph{Phys. Rev.
  Lett.} {\bf 110} (2013) 121101}, [\href{https://arxiv.org/abs/1210.0647}{{\tt
  1210.0647}}].

\bibitem{Sonbas:2014jya}
E.~Sonbas, K.~S. Dhuga, P.~Veres, G.~A. MacLachlan, F.~Dolek, T.~N. Ukwatta
  et~al., \emph{{Gamma-Ray Bursts: Temporal Scales and the Bulk Lorentz
  Factor}},
  \href{http://dx.doi.org/10.1088/0004-637X/805/2/86}{\emph{Astrophys. J.} {\bf
  805} (2015) 86}, [\href{https://arxiv.org/abs/1408.3042}{{\tt 1408.3042}}].

\bibitem{Gupta:2006jm}
N.~Gupta and B.~Zhang, \emph{{Neutrino Spectra from Low and High Luminosity
  Populations of Gamma Ray Bursts}},
  \href{http://dx.doi.org/10.1016/j.astropartphys.2007.01.004}{\emph{Astropart.
  Phys.} {\bf 27} (2007) 386--391},
  [\href{https://arxiv.org/abs/astro-ph/0606744}{{\tt astro-ph/0606744}}].

\bibitem{Campana:2006qe}
S.~Campana et~al., \emph{{The shock break-out of grb 060218/sn 2006aj}},
  \href{http://dx.doi.org/10.1038/nature04892}{\emph{Nature} {\bf 442} (2006)
  1008--1010}, [\href{https://arxiv.org/abs/astro-ph/0603279}{{\tt
  astro-ph/0603279}}].

\bibitem{Yuksel:2008cu}
H.~Yuksel, M.~D. Kistler, J.~F. Beacom and A.~M. Hopkins, \emph{{Revealing the
  High-Redshift Star Formation Rate with Gamma-Ray Bursts}},
  \href{http://dx.doi.org/10.1086/591449}{\emph{Astrophys. J.} {\bf 683} (2008)
  L5--L8}, [\href{https://arxiv.org/abs/0804.4008}{{\tt 0804.4008}}].

\bibitem{Strolger:2015kra}
L.-G. Strolger, T.~Dahlen, S.~A. Rodney, O.~Graur, A.~G. Riess, C.~McCully
  et~al., \emph{{The Rate of Core Collapse Supernovae to Redshift 2.5 From The
  CANDELS and CLASH Supernova Surveys}},
  \href{http://dx.doi.org/10.1088/0004-637X/813/2/93}{\emph{Astrophys. J.} {\bf
  813} (2015) 93}, [\href{https://arxiv.org/abs/1509.06574}{{\tt 1509.06574}}].

\bibitem{Wanderman:2009es}
D.~Wanderman and T.~Piran, \emph{{The luminosity function and the rate of
  Swift's Gamma Ray Bursts}},
  \href{http://dx.doi.org/10.1111/j.1365-2966.2010.16787.x}{\emph{Mon. Not.
  Roy. Astron. Soc.} {\bf 406} (2010) 1944--1958},
  [\href{https://arxiv.org/abs/0912.0709}{{\tt 0912.0709}}].

\bibitem{Dahlen:2004km}
T.~Dahlen et~al., \emph{{High redshift supernova rates}},
  \href{http://dx.doi.org/10.1086/422899}{\emph{Astrophys. J.} {\bf 613} (2004)
  189--199}, [\href{https://arxiv.org/abs/astro-ph/0406547}{{\tt
  astro-ph/0406547}}].

\bibitem{Schnabel:2013kma}
{\scshape ANTARES} collaboration, J.~Schnabel, \emph{{Muon energy
  reconstruction in the ANTARES detector}},
  \href{http://dx.doi.org/10.1016/j.nima.2012.12.109}{\emph{Nucl. Instrum.
  Meth.} {\bf A725} (2013) 106--109}.

\bibitem{Adrian-Martinez:2016fdl}
{\scshape KM3Net} collaboration, S.~Adrian-Martinez et~al., \emph{{Letter of
  intent for KM3NeT 2.0}},
  \href{http://dx.doi.org/10.1088/0954-3899/43/8/084001}{\emph{J. Phys.} {\bf
  G43} (2016) 084001}, [\href{https://arxiv.org/abs/1601.07459}{{\tt
  1601.07459}}].

\bibitem{Ackermann:2017pja}
{\scshape IceCube Gen2} collaboration, M.~Ackermann et~al., \emph{{The IceCube
  Neutrino Observatory}},  in \emph{{Proceedings, 35th International Cosmic Ray
  Conference (ICRC 2017): Bexco, Busan, Korea, July 12-20, 2017}}, 2017.
\newblock \href{https://arxiv.org/abs/1710.01207}{{\tt 1710.01207}}.

\end{thebibliography}\endgroup

\end{document}